\documentclass[aps,prb,twocolumn,groupedaddress,showpacs,10pt]{revtex4-1}
\usepackage{graphicx}
\usepackage{color}
\usepackage{calc}
\usepackage{array}
\usepackage{graphicx}
\usepackage{amsmath, amssymb}
\usepackage{natbib}
\usepackage{hyperref}
\usepackage{subfigure}
\usepackage{overpic}
\usepackage{tikz}

\newcommand{\red}[1]{{\color{black}{#1}}}

\begin{document}

\title{Topological superconductivity from first-principles II: \\
Effects from manipulation of spin spirals $-$ Topological fragmentation, braiding, and Quasi-Majorana Bound States}

\author{Andr\'as L\'aszl\'offy$^{1}$}
\author{Bendeg\'uz Ny\'ari$^2$} 
\author{G\'abor Csire$^{3,4}$}
\author{L\'aszl\'o Szunyogh$^{2,5}$}
\author{Bal\'azs \'Ujfalussy$^{1}$}

\affiliation{$^1$Wigner Research Centre for Physics, Institute for Solid State Physics and Optics, H-1525 Budapest, Hungary}

\affiliation{$^2$Department of Theoretical Physics, Institute of Physics, Budapest University of Technology and Economics, M\H uegyetem rkp.~3., HU-1111 Budapest, Hungary}

\affiliation{$^3$Materials Center Leoben Forschung GmbH, Roseggerstraße 12, 8700 Leoben, Austria.}
\affiliation{$^4$Catalan Institute of Nanoscience and Nanotechnology (ICN2), CSIC, BIST, Campus UAB, Bellaterra, Barcelona, 08193, Spain}

\affiliation{$^5$ELKH-BME Condensed Matter Research Group, Budapest University of Technology and Economics, M\H uegyetem rkp.~3., HU-1111 Budapest, Hungary}

\date{\today}

\begin{abstract}

Recent advances in electron spin resonance techniques
have allowed the manipulation of the spin of individual atoms,
making magnetic atomic chains on superconducting hosts \red{one of the most} promising platform
where topological superconductivity can be engineered.
Motivated by this progress, we provide a detailed, quantitative description of the effects
of manipulating spins in realistic nanowires by applying a
first-principles-based computational approach
to a recent experiment: an iron chain deposited on top of Au/Nb heterostructure.
As a continuation of \red{the preceding paper [B. Ny\'ari \textit{et al.} Phys. Rev. B (2023)]},
experimentally relevant computational experiments are performed in spin spiral chains
that shed light on several concerns about practical applications and add new aspects to the interpretation of recent experiments.
We explore the stability of topological zero energy states,
the formation and distinction of topologically trivial and non-trivial
zero energy edge states, the effect of local changes in the exchange fields,
the emergence of topological fragmentation,
and the shift of Majorana Zero Modes along the superconducting nanowires 
opening avenues toward the implementation of a braiding operation.

\end{abstract}

\maketitle

\section{Introduction}

This is the second in a series of two papers to describe the properties of magnetic atomic chains deposited on $s$-wave superconductors from first principles. Our research is motivated by the intense race to detect Majorana zero modes (MZMs), 
that could provide a unique platform for quantum computing\cite{Alicea2012, Beenakker2020, Marra2022review} free from significant decoherence.
However, these efforts are severely hampered \red{by the following problems.} 
(i) The essential topological properties of MZMs are still not
unambiguously confirmed in experiments. In fact, topologically trivial
zero energy states can still bear many features of MZMs, such as
localization with spin-polarized characteristics\cite{Jeon2017, Kster2022} or the fractional Josephson-effect\cite{Chiu2019}.
(ii) Possible mechanisms that may destroy topological superconductivity in nanowires,
have to be identified to guide experiments toward the realization of robust MZMs.
(iii) Basic elements of braiding protocols need to be established that enable the construction of more sophisticated quantum gates.
\red{Minimal tight-binding models}\cite{kitaevchain, Fu2008, Lutchyn2010, oreg2010helical, NadjPerge2013, Hell2017}
may provide some helpful guidance in approaching these problems, such as
pinpointing possible physical mechanisms that may cause topologically trivial zero energy states and
those which may destroy MZMs. However, they are incapable to describe and capture the details
of a particular experimental realization. Instead, an accurate first principles model is needed
to distinguish which mechanism might be at play and describe it in a quantitative manner.

In the preceding paper\cite{part1}, henceforth referred to as paper I, we made an attempt to answer questions related to point (i). We showed that a minigap and zero energy peaks, so-called Majorana Zero Modes - can be observed in the local density of states (LDOS) within the superconducting gap of the host. We also showed, that the MZM is localized to the edge atoms of the chain, and discussed its spatial dispersion. We examined the nature of the MZM edge states based on the triplet and singlet order parameters and found the state to be an internally antisymmetric triplet (IAT). Based on the antisymmetric properties of the Bogoliubov-deGennes (BdG) equations, we also developed two different quantities, the quasiparticle charge density of states (CDOS) and the singlet superconducting order parameter to signal the topological nature of the minigap.
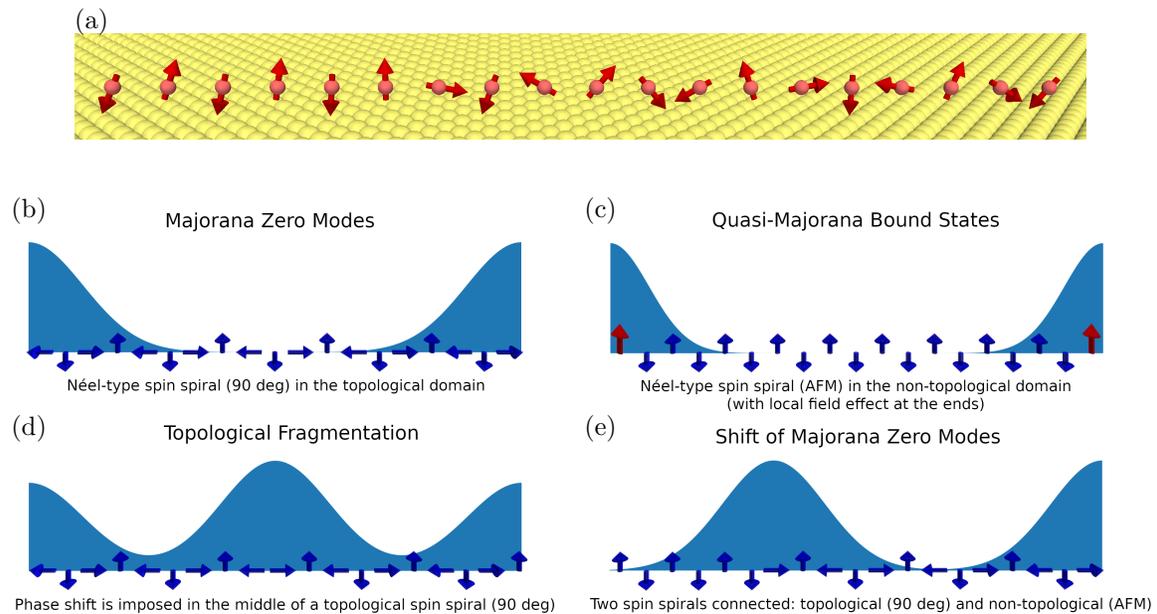
\begin{figure*}
\begin{overpic}[width=0.75\textwidth]{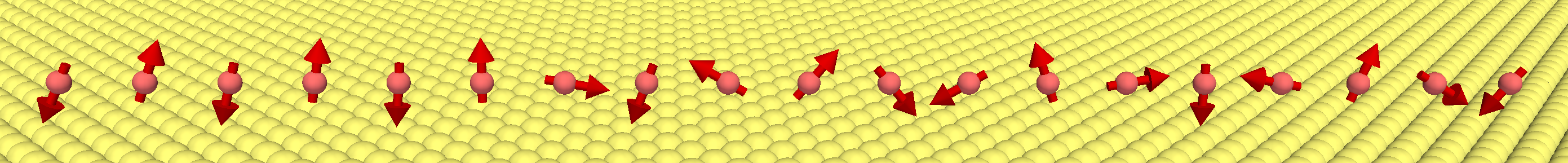}
\put (0,11) {\normalsize (a)}
\end{overpic}
\vspace{1cm}\\
\begin{overpic}[width=0.85\textwidth]{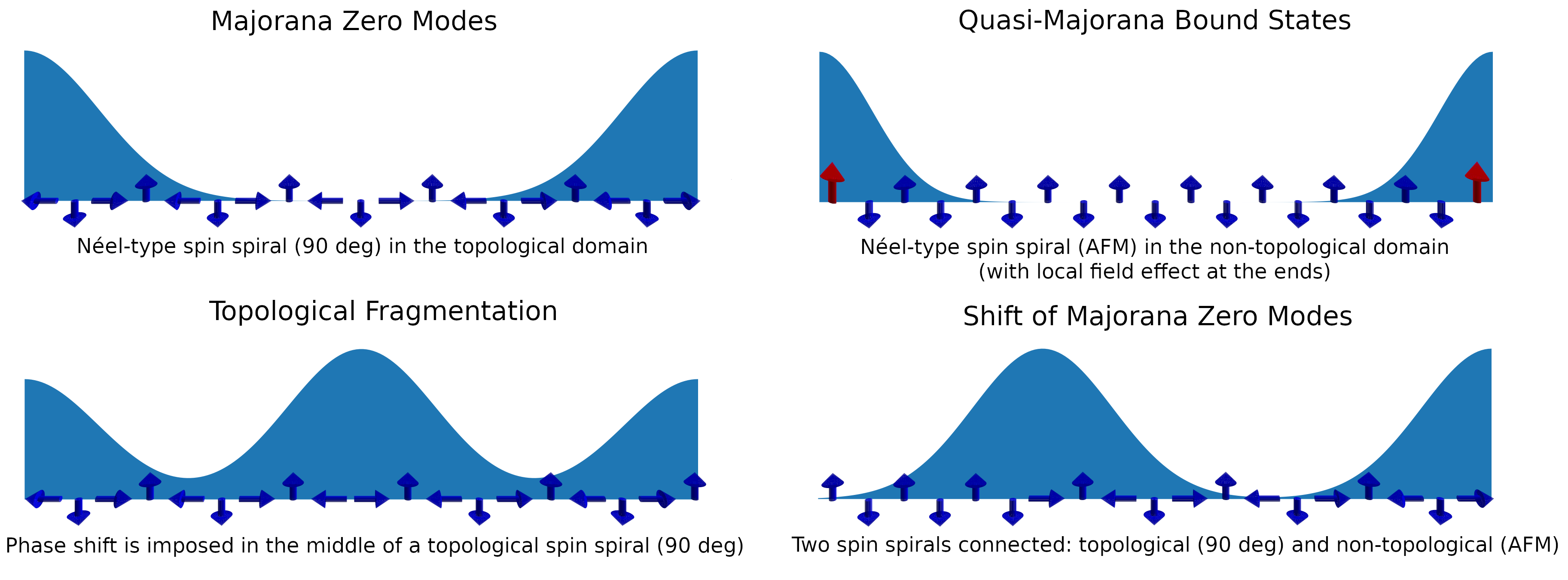}
\put (0,36) {\normalsize (b)}
\put (0,17) {\normalsize (d)}
\put (50,36) {\normalsize (c)}
\put (50,17) {\normalsize (e)}
\end{overpic}
\caption{\label{fig:schematic} Schematic figures about the observed pool of phenonema.
(a) The illustration of the 19 atomic 2a-[100] Fe chain on Nb(110) covered with a single monolayer Au. The spin configuration presents a case that has been thoroughly investigated: a $175^\circ$ Néel-type spiral (left 6 atoms) connected to a $100^\circ$ spiral (right 14 atoms), the sixth atom is an element of both spirals.
(b) Majorana Zero Modes at the ends of the topological superconducting nanowire,
(c) Quasi-Majorana Bound states arising from local field effect on non-topological spin spiral states at the ends of the nanowire,
(d) Topological Fragmentation: due to a phase shift on topological spin spiral states in the middle of the nanowire
additional MZMs appear,
(e) The shift of MZM due to the combination of two topological spin spirals with different spiraling angles.
}
\end{figure*}
In the current paper, we make an attempt to answer questions related to points (ii) and (iii) via a series of computational experiments on various spin spiral chains: the combination of spin spirals with different spiraling angles and different phases,
and explore the effects of local changes in the spin directions and exchange field in realistic nanowires.
Our main focus shall still be the same system that we used as a model in our calculations so far, and is recently under 
experimental investigations\cite{beck2023search}:
a 19-atom-long Fe chain with 2a (a=330 pm) nearest neighbor distance in [100] direction
(in short: 2a-[110] Fe$_{19}$ chain),
placed on the (110) surface of an epitaxial Au monolayer covering the surface of Nb(110)
as illustrated in Fig.~\ref{fig:schematic}a.

In Section~\ref{sec:MZM} to test the stability and robustness of MZMs we perform additional computational experiments applying random perturbations to the orientations of many spirals which show signatures of robust topological superconductivity and MZMs.
In Section~\ref{sec:phase} we \red{discuss} 
the concept of topological fragmentation
that can cause the appearance of several MZMs in the internal region of the nanowire (see also Appendix~A). These emergent MZMs can hybridize with the MZMs at the edges developing into finite energy Yu-Shiba-Rusinov (YSR)
states\cite{yu1965bound,shiba1968classical,rusinov1969superconductivity,Balatsky2006}.
By imposing phase differences in the spin spirals as illustrated in Fig.~\ref{fig:schematic}d, we shall determine a range of phase shifts that may cause topological fragmentation and, therefore,
the destruction of MZMs in short nanowires. 
\red{This scenario introduces a domain wall into the nanowire which was also investigated
on the model level in Refs.~\onlinecite{Flensberg2010, marra2017controlling, Marra2022}.
The idea of topological fragmentation is explained in a rather general tight-binding model as well in Appendix~A.}
Quite on the practical side,
the combination of two spin spirals with different spiraling angles, one from the non-topological and
the other from the topological domain (see Fig.~\ref{fig:schematic}e for illustration),
allows to shift MZMs in superconducting nanowires as it is presented in Section~\ref{sec:shift}.
This may serve as a basic element of a braiding operation in theory. It should be noted that the recent advances in experimental technology gives hope to manipulate the local atomic-scale magnetic structure (and realize the computer experiment)  using a combination of scanning tunneling
microscopy and electron-spin resonance techniques\cite{Yang2019,Willke2021,Phark2022}.
A similar study can also be carried out in the non-topological domain presented in Section~\ref{sec:QMBS} where we do show how trivial zero-energy edge states, called Quasi-Majorana Bound States (QMBS), may arise due to local field effects (Fig.~\ref{fig:schematic}c). We also show, how they can be distinguished from the long-sought MZMs (Fig.~\ref{fig:schematic}b) within a first-principles calculation in the superconducting state. 
We summarize our main findings in Section~\ref{sec:summary} while the technical details are provided in Appendix~B.
\begin{figure*}
\centering 
	\includegraphics[scale=1]{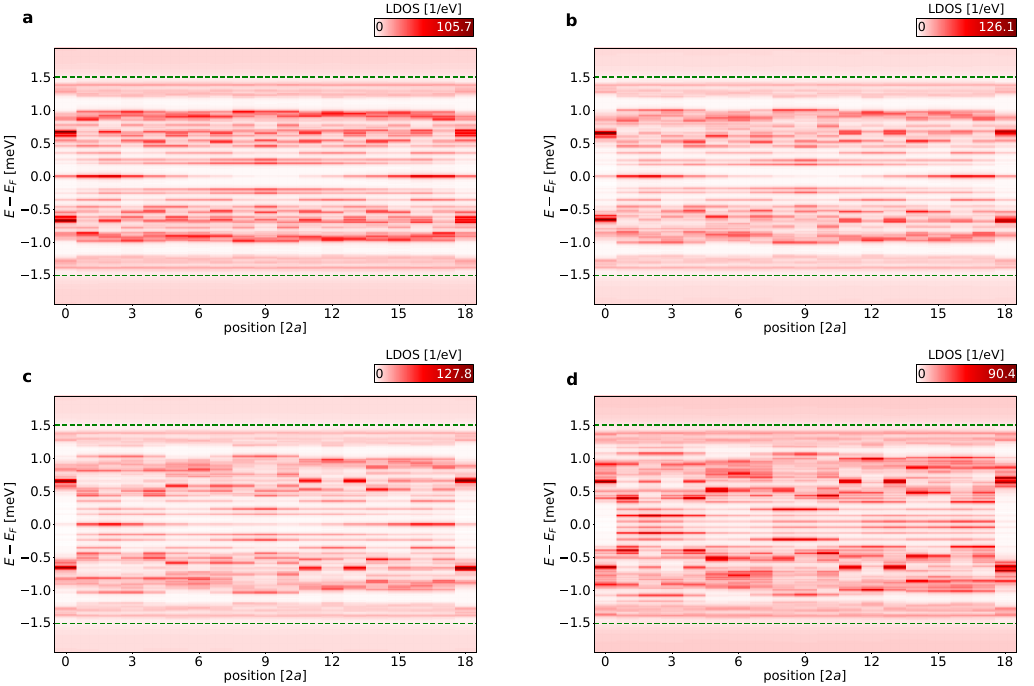}
 	\caption{\label{fig:supp-noise} Stability of zero energy edge states against white noise. 
  \red{The directions of magnetic moments in a 100$^\circ$ spiral are perturbed on each site by a randomly generated angle. The random angle was allowed to vary  between 0$^\circ$ and (a) 60$^\circ$, (b) 90$^\circ$, (c) 120$^\circ$ and (d) 172$^\circ$}
  .}
\end{figure*}

While in the previous paper, we utilized full charge self-consistency in the normal state, in order to be able to identify Majorana Zero Modes most unambiguously, with the understanding obtained from those calculations we now step away somewhat from that requirement. In order to reduce the cost of the calculations, we will calculate self-consistent potentials for the impurity system only in the ferromagnetic (normal) state.  In the subsequent calculations for various spin spirals in the superconducting state, we shall adopt a frozen potential picture, and rotate the magnetic moments without further self-consistency. Differences caused by this approximation compared to the previous, fully self-consistent calculations are also interesting from a purely methodological point of view and will be pointed out where appropriate.

\begin{figure*}
 \centering
    \begin{overpic}[width=0.45\textwidth]{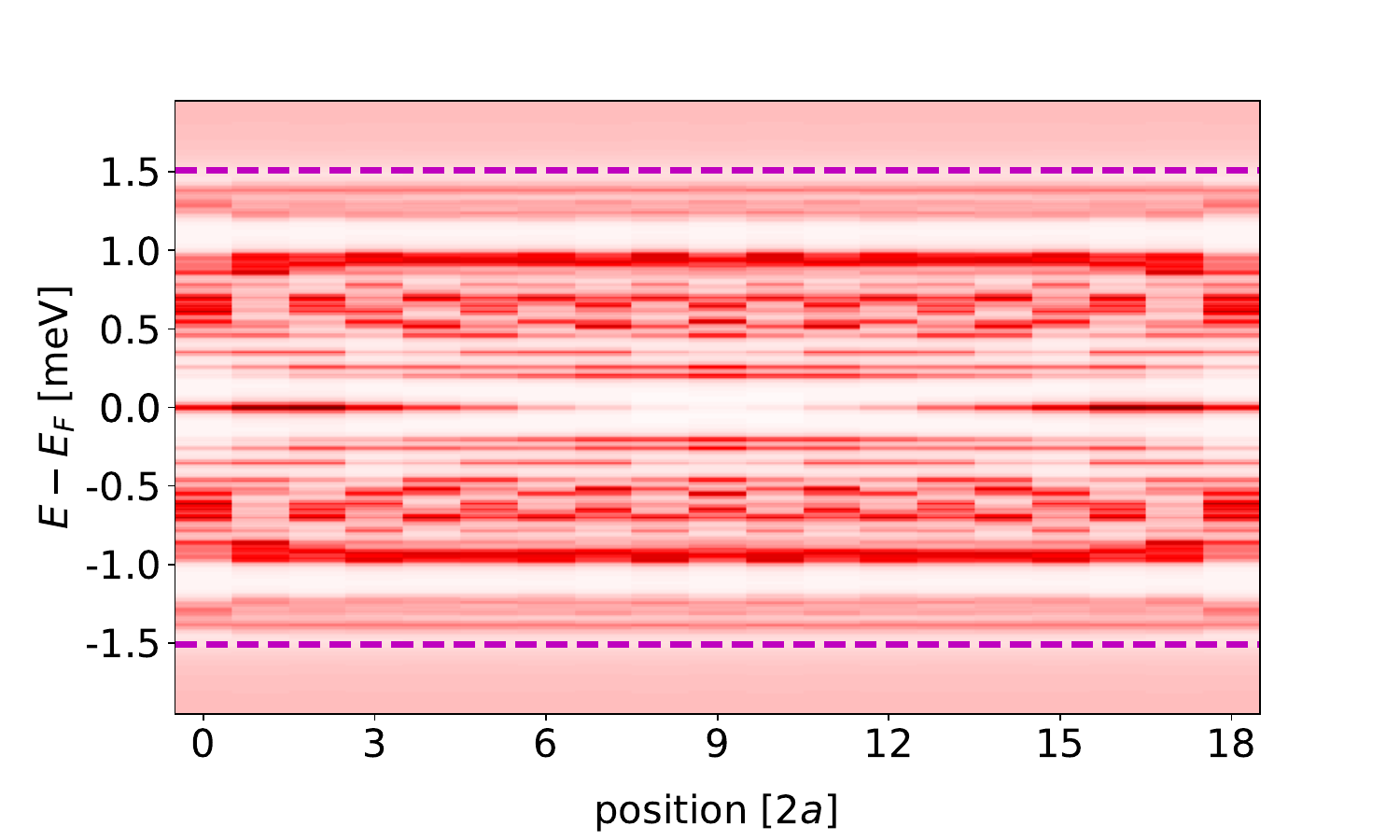}
    \put (0,50) {\normalsize (a)}
    \end{overpic}
    \begin{overpic}[width=0.45\textwidth]{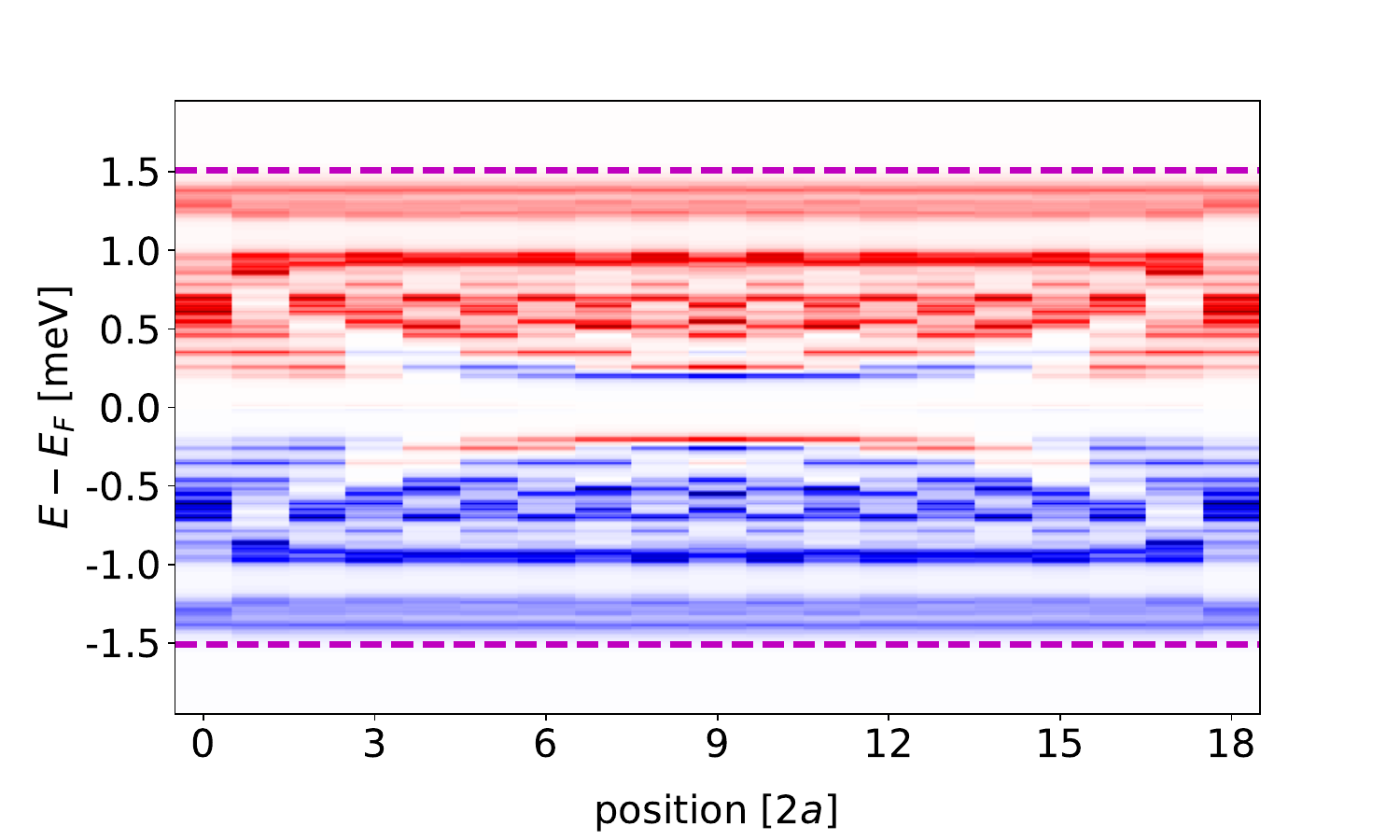}
    \put (0,50) {\normalsize (b)}
    \end{overpic}\\
    \begin{overpic}[width=0.45\textwidth]
{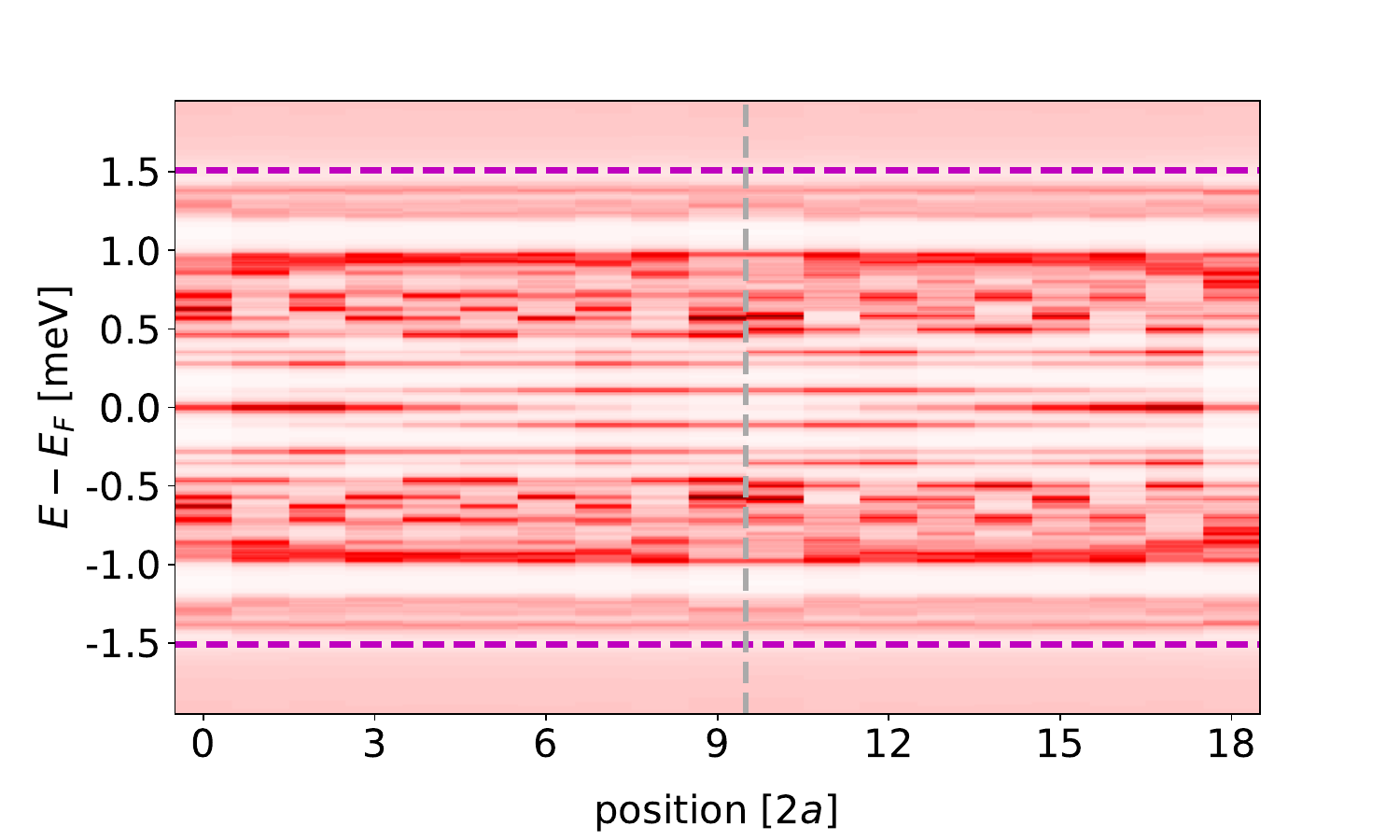}
	\put (0,50) {\normalsize (c)}
    \end{overpic}
    \begin{overpic}[width=0.45\textwidth]{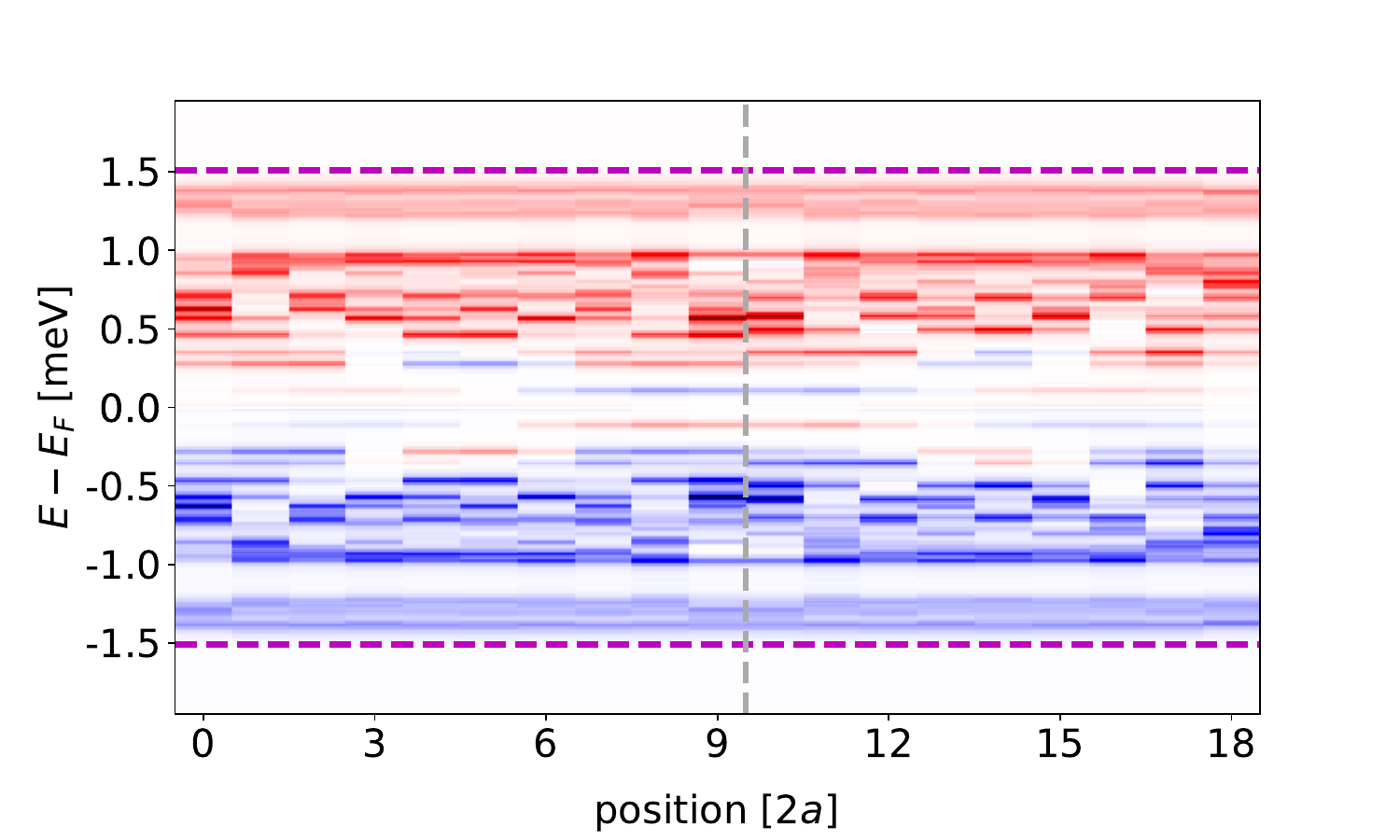}
        \put (0,50) {\normalsize (d)}
    \end{overpic}\\
    \begin{overpic}[width=0.45\textwidth]{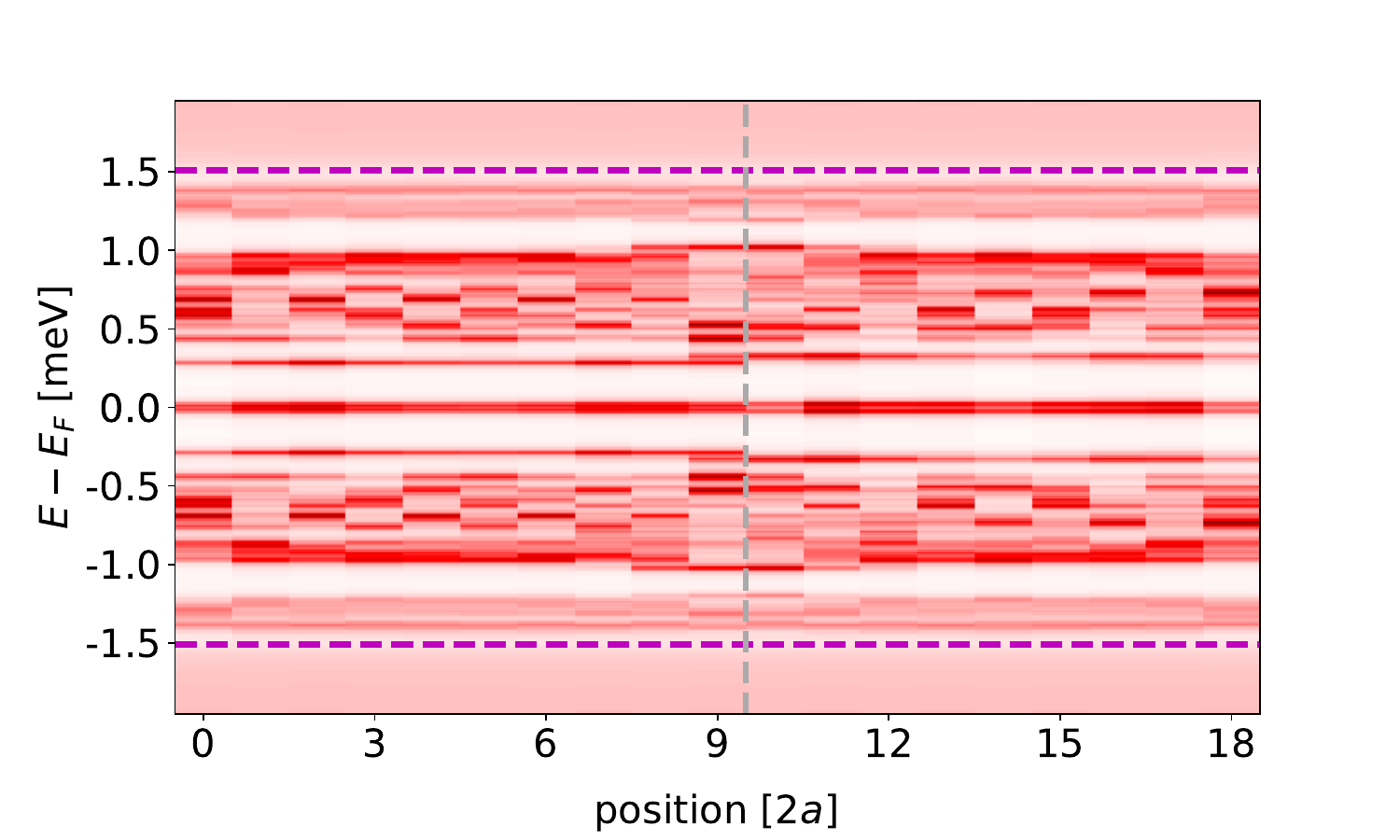} 
	\put (0,50) {\normalsize (e)}
    \end{overpic}
    \begin{overpic}[width=0.45\textwidth]{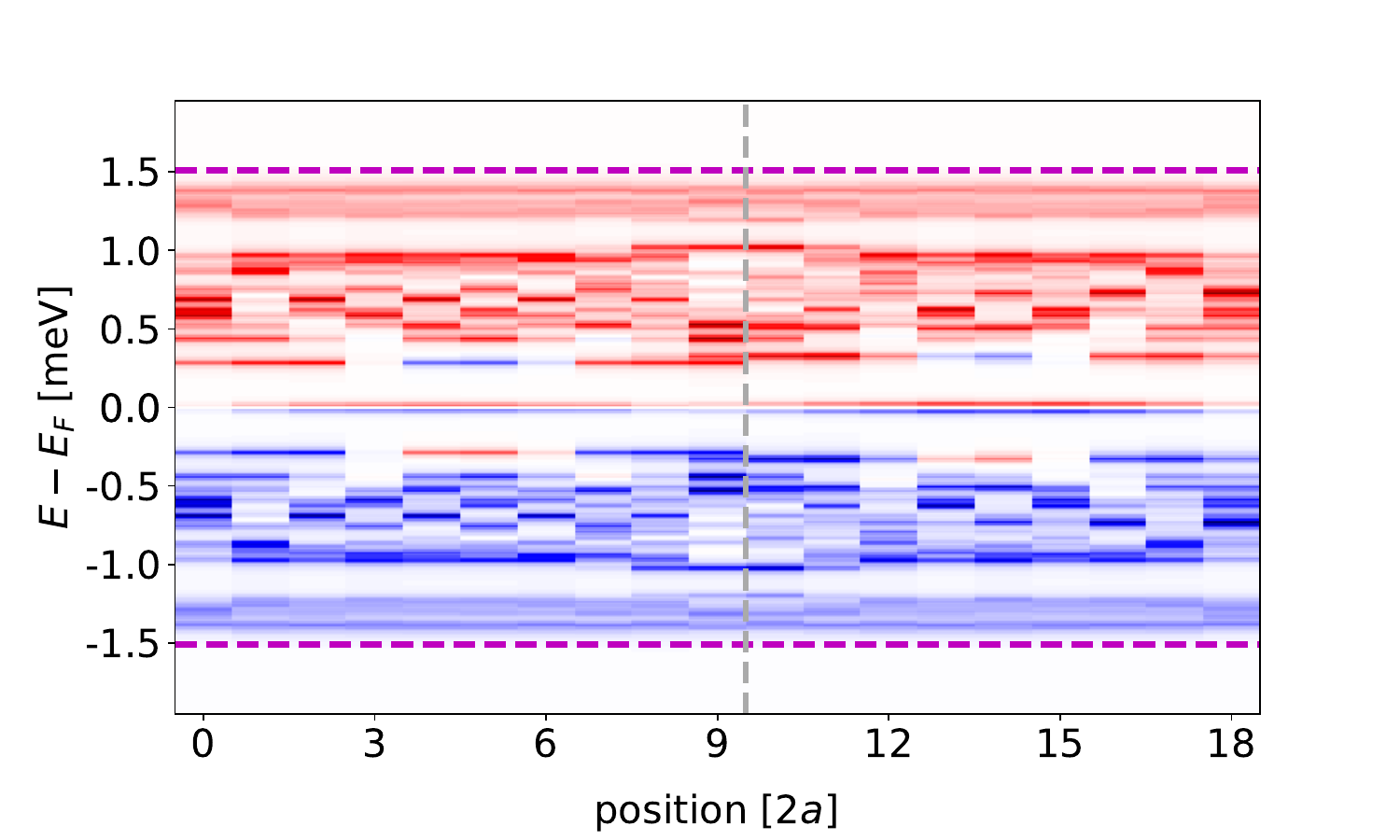}
     \put (0,50) {\normalsize (f)}
    \end{overpic}\\
    \begin{overpic}[width=0.45\textwidth]{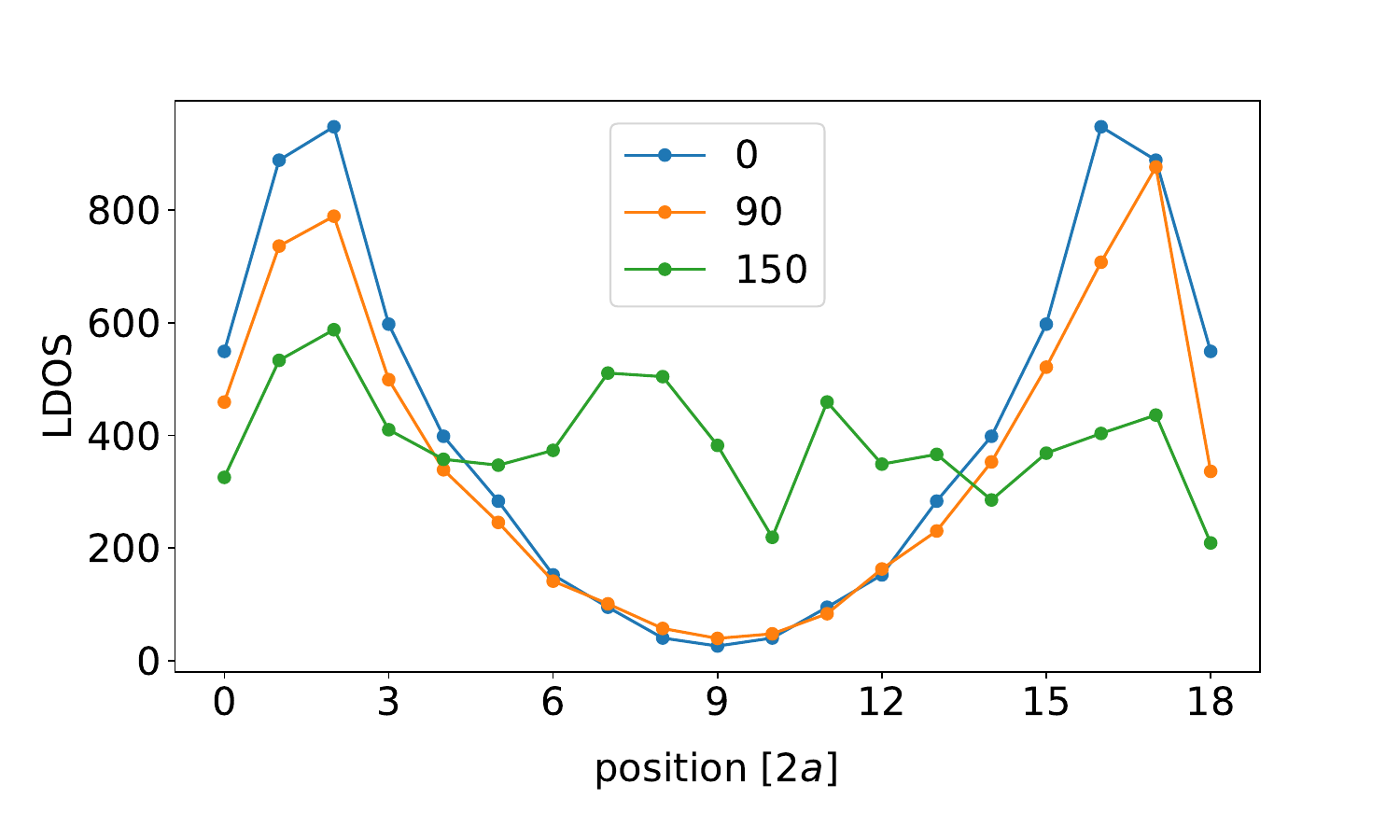}
    \put (0,50) {\normalsize (g)}
    \end{overpic}
	\caption{\label{fig:local-phase} Local behavior of the combination of two Néel-type spin spirals with the same spiraling angles (100$^\circ$ spiral) but linked with an additional phase shift applied after site 9.
    Local DOS without any additional phase (original 100$^\circ$ spin spiral) (a),
    and with the additional phase of 90$^\circ$(c)  and 150$^\circ$ (e).
    Subfigures (b), (d), and (f) are the corresponding Local CDOS plots.
    (g) Local DOS at the Fermi level representing the localization length in these cases.
    }
\end{figure*}

\section{Stability of MZM-s against random perturbations in the spin directions \label{sec:MZM}}
In an attempt to model the robustness of the MZMs against fluctuations of the atomic moments, we applied random perturbations to the orientations of many spirals. Such a calculation is a primitive way to model the effect of thermal fluctuations. These calculations resulted in a rather extensive dataset, which we can only sample here. 
Technically, we took several spirals and applied a random perturbation to the direction of the magnetization on each lattice site. The value of the maximum allowed perturbation was varied between 8-172 degrees, meaning that we used the $0.005-0.995$ part of Bloch's sphere around the original spin direction to generate the new, perturbed direction. 

The results for the perturbed 100$^\circ$ spirals are shown in Figure \ref{fig:supp-noise}. We used the SCF potentials of the unperturbed 100$^\circ$ spiral and turned off the exchange field at non-magnetic sites so that the SCF calculations did not have to be repeated. The maximum perturbation angle in the Figure was 60$^\circ$, 90$^\circ$, 120$^\circ$, or 172$^\circ$, corresponding to the noise where the new spin direction is selected from the 0.25, 0.5, 0.75, or 0.995 part of the unit sphere surface, respectively. Our main result, as can be seen in the figure is, that the overall features of the LDOS, and especially the positions of the \red{zero energy peaks (ZEPs)} are unchanged compared to the unperturbed case. For small perturbations (where the maximum angle is below 60$^\circ$) we saw very little change in the LDOS itself.  For higher values of the maximum angle, as can be seen in the figure, we obtained a smaller \red{but still topological} minigap, which kept decreasing as the maximum angle of the perturbation increased. Around a value of 120$^\circ$, the minigap starts to get filled, but the edge states at zero energy still remain virtually unaffected. In the case of the almost totally random spin directions (with 172$^\circ$ maximum perturbation), both the minigap and the \red{ZEP} finally disappear. Based on a larger dataset of our calculations, we believe that this stability can be attributed to the rather large range of spiraling angles, where \red{ZEP}s can be found. Since the \red{ZEP} remains unaffected by such a surprisingly large range of perturbation of the local directions, we conclude that it is robust against such fluctuations, which in reality can be caused by temperature. 

\red{It is worthwhile to mention that the presence of quantum dots, inhomogeneous potential, random disorder in the chemical potential,
random fluctuations in the superconducting gap was considered for a semiconductor-superconductor hybrid system model
in Ref.~\onlinecite{Pan2020}. Their results suggested that the MZMs are immune to weak disorder, however, strong disorder can
completely suppress topological superconductivity and can lead to trivial ZEPs.
Our findings for the random perturbations in the spin directions of spin-spiral nanowires show similar qualitative behavior: MZMs are immune to this type of fluctuations
but up to surprisingly strong disorder. In the case of extremely strong disorder, peaks that are very close to zero energy may arise and can mimic the zero energy feature of MZMs.}

\section{Phase shifts and topological fragmentation   \label{sec:phase}}
In long spirals, several spin spiral fragments can freeze-in in smaller parts of the chain\cite{Choi2019}, leading to phase differences in spiraling angles between them. By spiral fragments, we mean a fragment of the chain where the spiraling angle is the same as everywhere in the chain, but there are jumps in the phase between fragments.  In order to investigate this scenario, especially its effect on the MZMs,
the impact of an additional phase shift on spin spirals (within the topological domain)
is investigated on the example of the 2a-[110] Fe$_{19}$ nanowire in Fig.~\ref{fig:local-phase}. Here the result of a computer experiment is presented,
where two spin spirals with the same (100$^\circ$) spiraling angles were connected, but with an additional phase shift between them at sites 9 and 10.
This creates an artificial boundary in the middle of the chain.
\red{In Ref.~\onlinecite{marra2017controlling} several other proposals were made how a domain wall can appear for MZMs in nanowire (e.g. with amplitude modulated fields) showing similar behavior. }
One should keep in mind that in the topological phase, Majorana operators from
different sites are paired together\cite{kitaevchain}
and such a boundary (domain wall) may cause topological fragmentation. This can be first looked at in the context of a simple tight-binding model, as the one outlined in Appendix~A. The model clearly shows the effect of topological fragmentation. 
Interestingly, this conclusion remains true in the realistic iron nanowire as well. We may observe, that for a certain range of the additional phase shift (like for example for 150$^\circ$), two MZMs appear at the Fermi energy
without interacting with each other in the middle of the nanowire (see Figure~\ref{fig:local-phase}e). However, these additional zero energy states hybridize with the edge states of the original spiral, forming two well-separated (shorter) chains between sites 0 and 9, and between sites 10 and 18 with slightly overlapping MZMs. Note that due to the overlap, they are slightly split around the Fermi energy as well. 
If the phase shift is increased to 180$^\circ$, the original MZMs at the edges of the chain are regained, while the states in the middle of the chain develop into finite energy YSR states.
\begin{figure*}
    \begin{overpic}[width=0.45\textwidth]{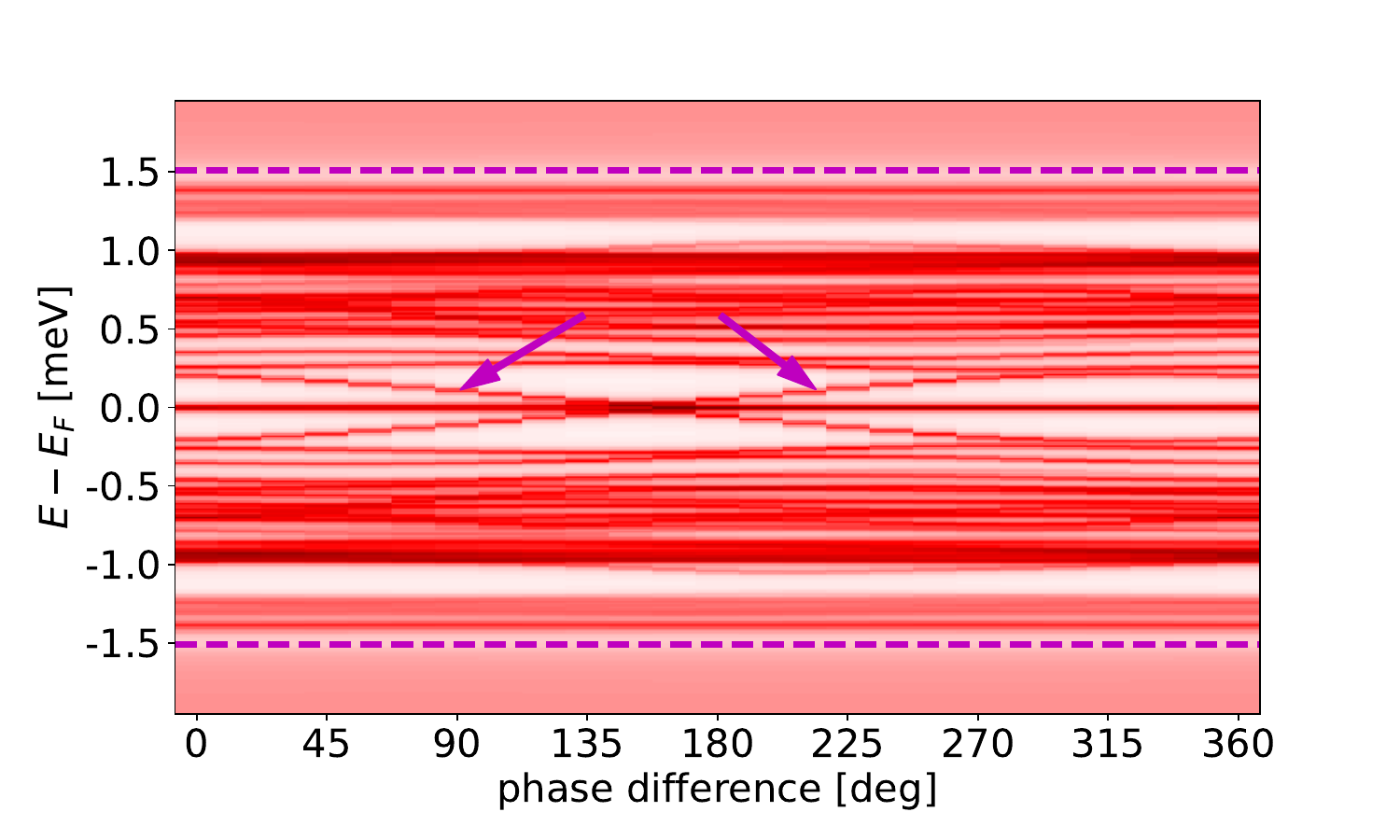}
    \put (0,50) {\normalsize (a)}
    \end{overpic}
    \begin{overpic}[width=0.45\textwidth]{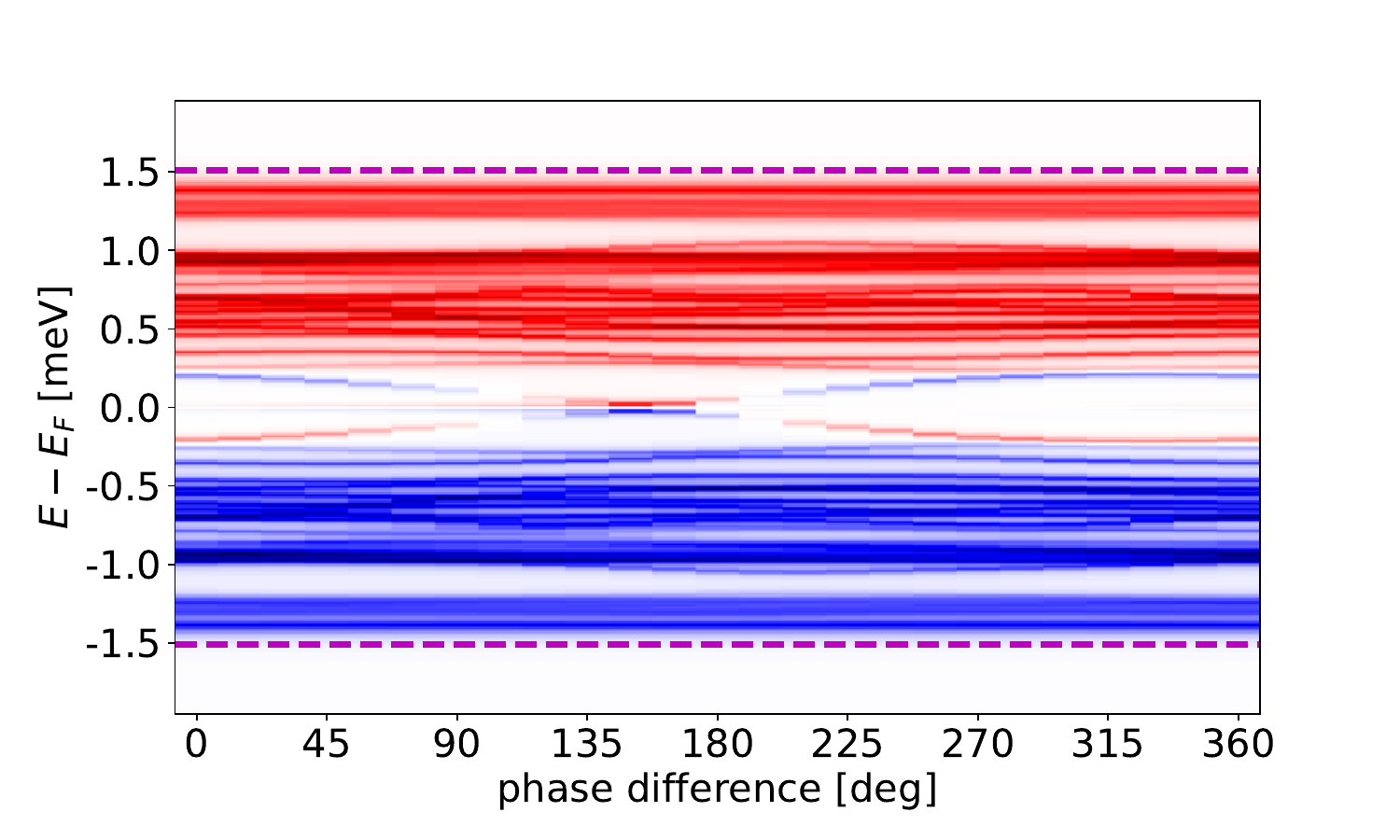}
    \put (0,50) {\normalsize (b)}
    \end{overpic}
\caption{\label{fig:phase_diff} Contact of Néel-type spin spirals with various phases.
The total DOS (a) and the CDOS (b) as a function of phase shift where two spirals have been linked with the same spiraling angles (100$^\circ$ spiral) but with various phase shifts between sites 9 and 10.
The arrows indicate the YSR states shifting between the edge of the minigap and the Fermi energy. Between about 135$^\circ$ and 180$^\circ$, these states evolve into additional MZMs in the middle of the chain.}
\end{figure*}
During the argumentation above, it still needs to be proven, that the emergent zero energy states in the middle are indeed MZMs and the nanowire is in a state of topological fragmentation. This conclusion can be drawn either from the plots of the local CDOS (LCDOS) or the energy-resolved singlet order parameter, as it was described in detail in paper I. We will now follow the argument based on the CDOS, although both quantities perfectly reflect the concept of topological fragmentation. As we showed in paper I, at both ends of a nanowire where MZMs appear, the LCDOS does not demonstrate band inversion, since it represents the boundary of two topologically different regions dissolving in the appearance of MZMs. However, on the sites in-between, the LCDOS should demonstrate the band inversion
as it can be observed in Fig.~\ref{fig:local-phase}b and d.
The introduction of a phase difference moves the energy of the inverted Shiba states away from the edge of the minigap, closer to zero energy, and pushes them further toward zero as the phase difference increases. This effect bears similarities with the shift of single impurity YSR states with the canting angle of the magnetic moment\cite{Kyungwha2023}.
This also leaves the remaining Shiba states at the minigap edge around the middle of the nanowire non-inverted (Fig.~\ref{fig:local-phase}f).
At some angles, the inverted Shiba states reach zero energy, that is when  MZMs appear in the middle of the chain since the LCDOS now exhibits exactly the same structure (both between sites 0-9, and 10-18 on Fig.~\ref{fig:local-phase}f) for these zero energy states as it does for MZMs at the ends of the full chain in Fig.~\ref{fig:local-phase}b. 
In Fig.~\ref{fig:local-phase}g we show the site-resolved Local DOS at the Fermi level for the imposed additional phase
of 90$^\circ$ and 150$^\circ$ together with the original case (without any additional phase).
One may observe that
for the additional phase of 150$^\circ$, where the MZMs appear in the middle of the iron chain as well, the hybridization with the original edge states destroys their localized behavior and makes them unfeasible for topological quantum computation at this size of the chain. Here we should mention that the LDOS plot for the spiral with $0^\circ$  additional phase (the original spiral) is slightly different compared to the corresponding figure in paper I, which is a result of the non-selfconsistent treatment - for reasons mentioned in the introduction - of the spirals in the present paper. 

Fig.~\ref{fig:phase_diff} summarizes the effect of various additional phases on a 100$^\circ$ spin spiral state of the iron nanowire.
The quasi-particle DOS shown in Fig.~\ref{fig:phase_diff}a demonstrates 
how the state at the top of the gap (marked with arrows) breaks off and moves to zero energy as a function of the additional phase (and moves back to the gap edge, as it is increased further).
It should be noted, that there is a surprisingly wide variety of additional phases causing topological fragmentation
between around 135$^\circ$ and 180$^\circ$, emphasizing the practical importance of this phenomenon as a danger to MZM-based qubits.
It can also be seen in Fig.~\ref{fig:phase_diff}b that the overall signature of band inversion vanishes at some phase differences on the CDOS plot which is a consequence of the rather short length of the iron nanowire in the calculation.

\red{In this section we have predicted the exact numerical details for phase biases which introduce a domain wall and topological fragmentation. 
These findings can be directly checked with recent experimental techniques of ESR-STM and once they are confirmed they also substantiate the topological behavior.}

\section{Spirals with different spiraling angles and braiding 
         \label{sec:shift}}

\begin{figure*}
	\centering
    \begin{overpic}[width=0.45\textwidth]{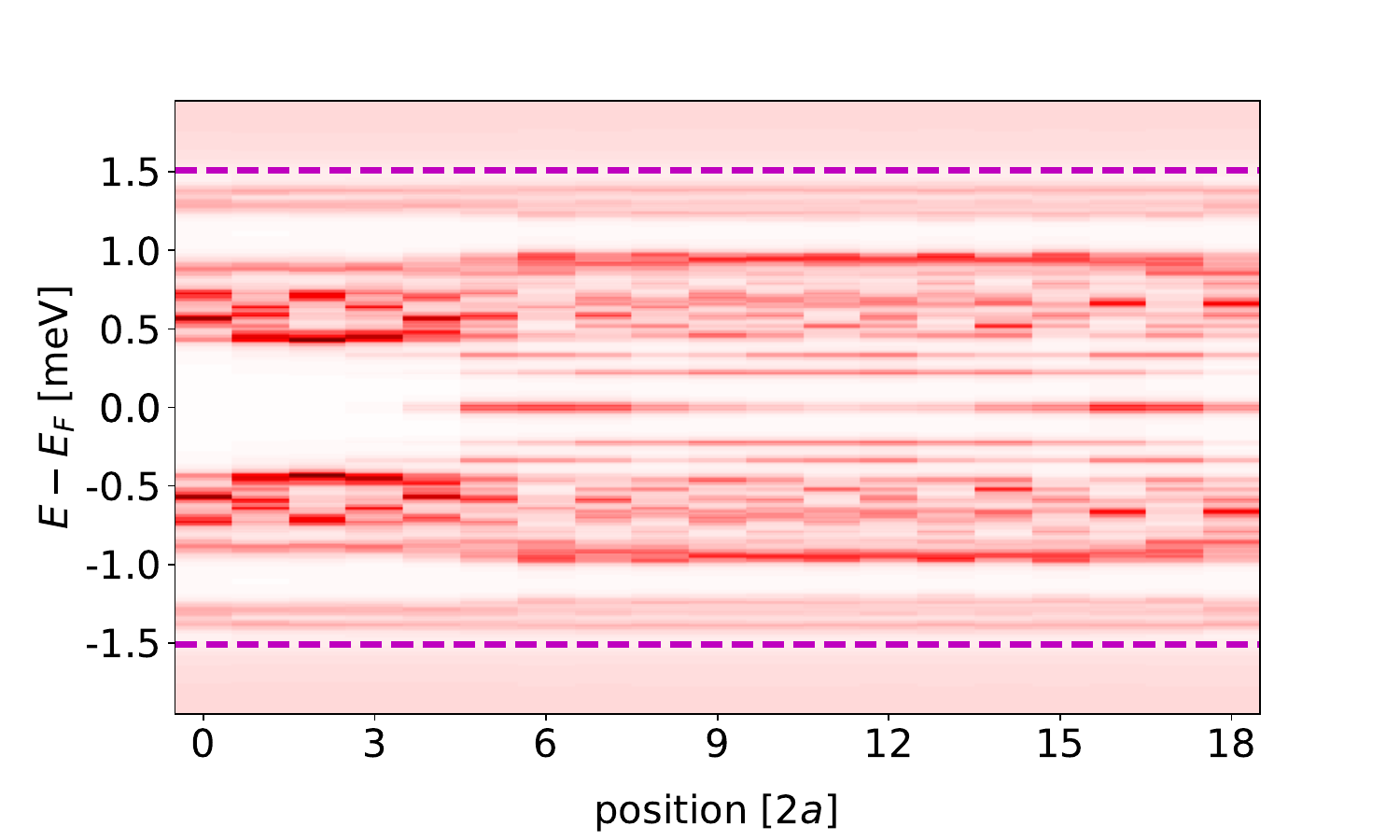}
    \put (0,50) {\normalsize (a)}
    \end{overpic}
    \begin{overpic}[width=0.45\textwidth]{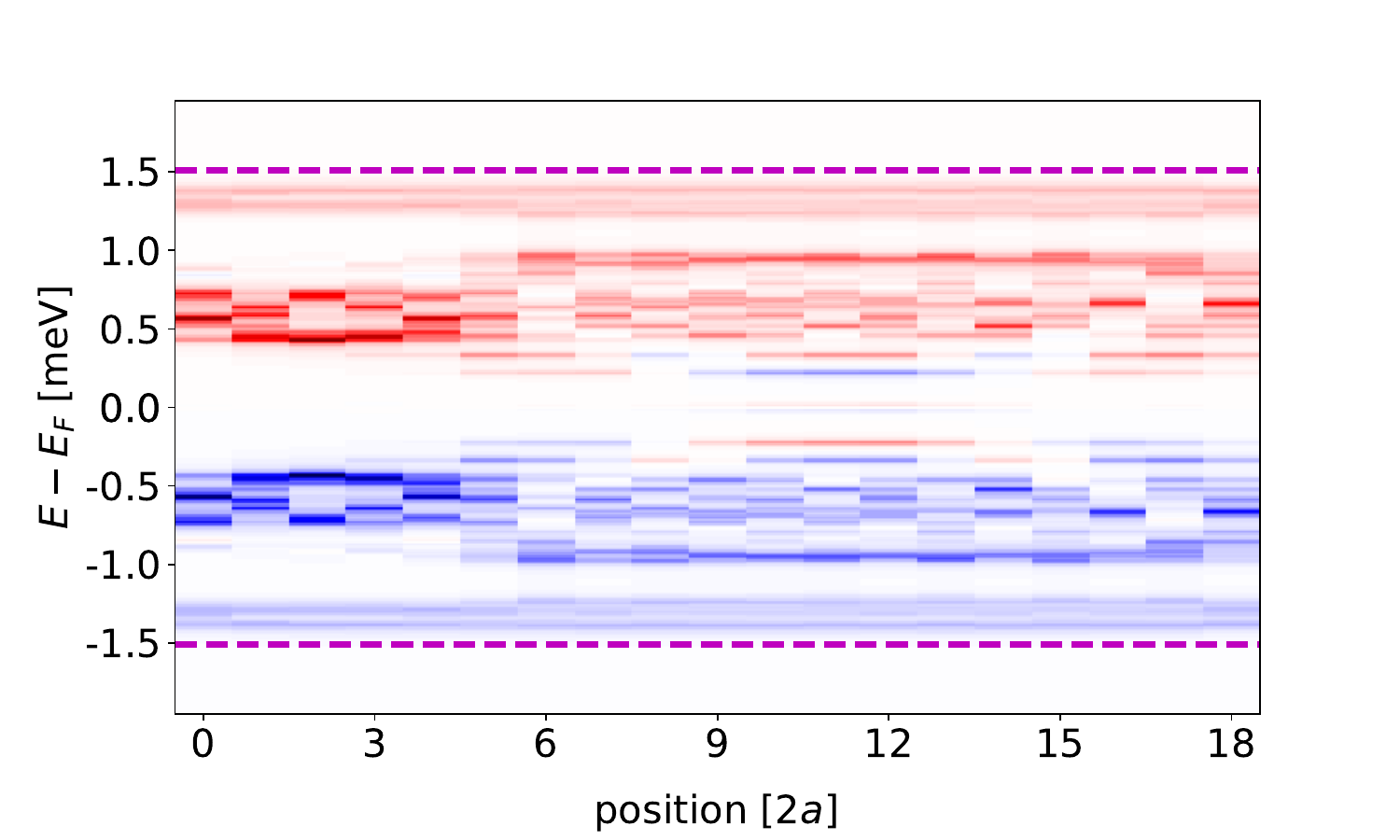}
    \put (0,50) {\normalsize (b)}
    \end{overpic}\\
    \begin{overpic}[width=0.45\textwidth]{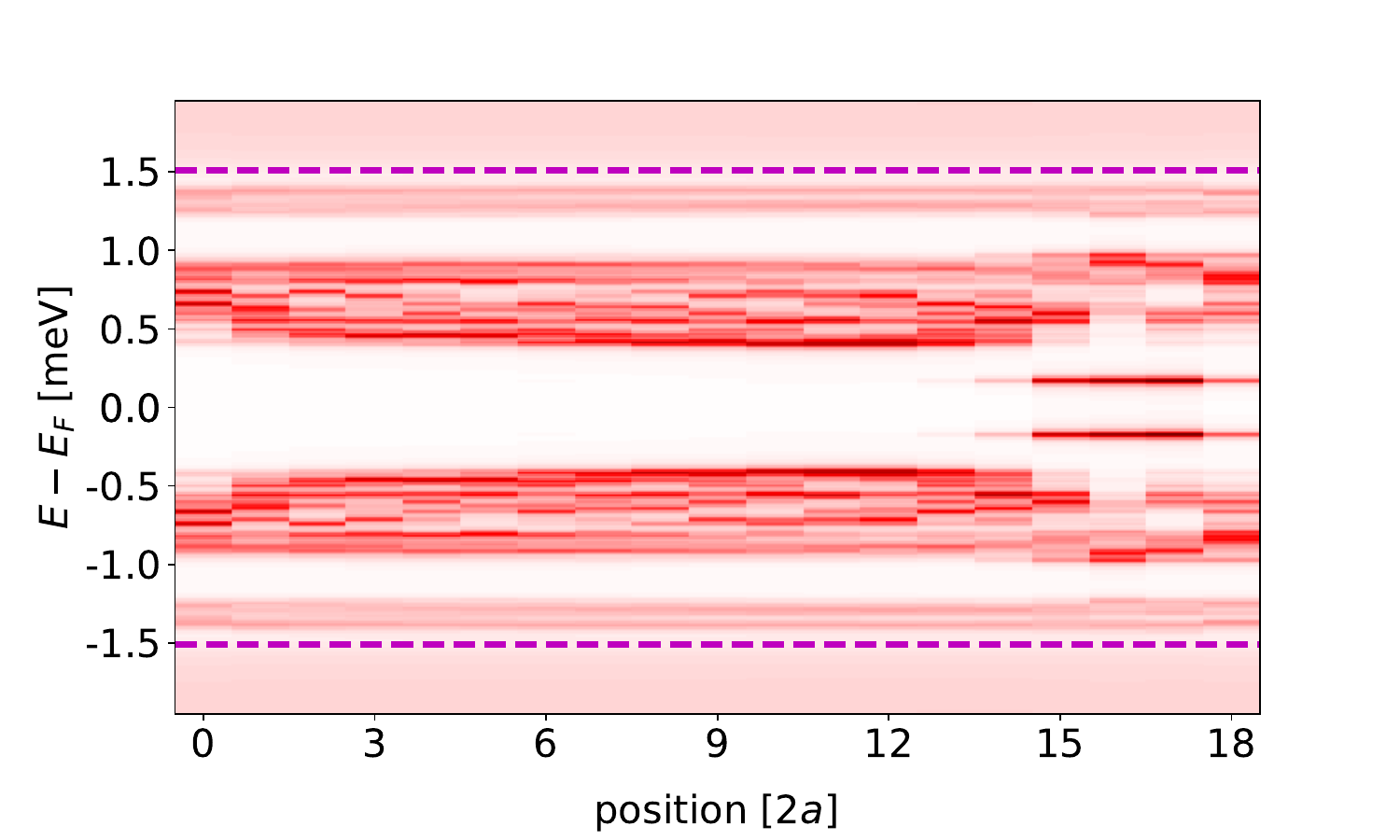}
    \put (0,50) {\normalsize (c)}
    \end{overpic}
    \begin{overpic}[width=0.45\textwidth]{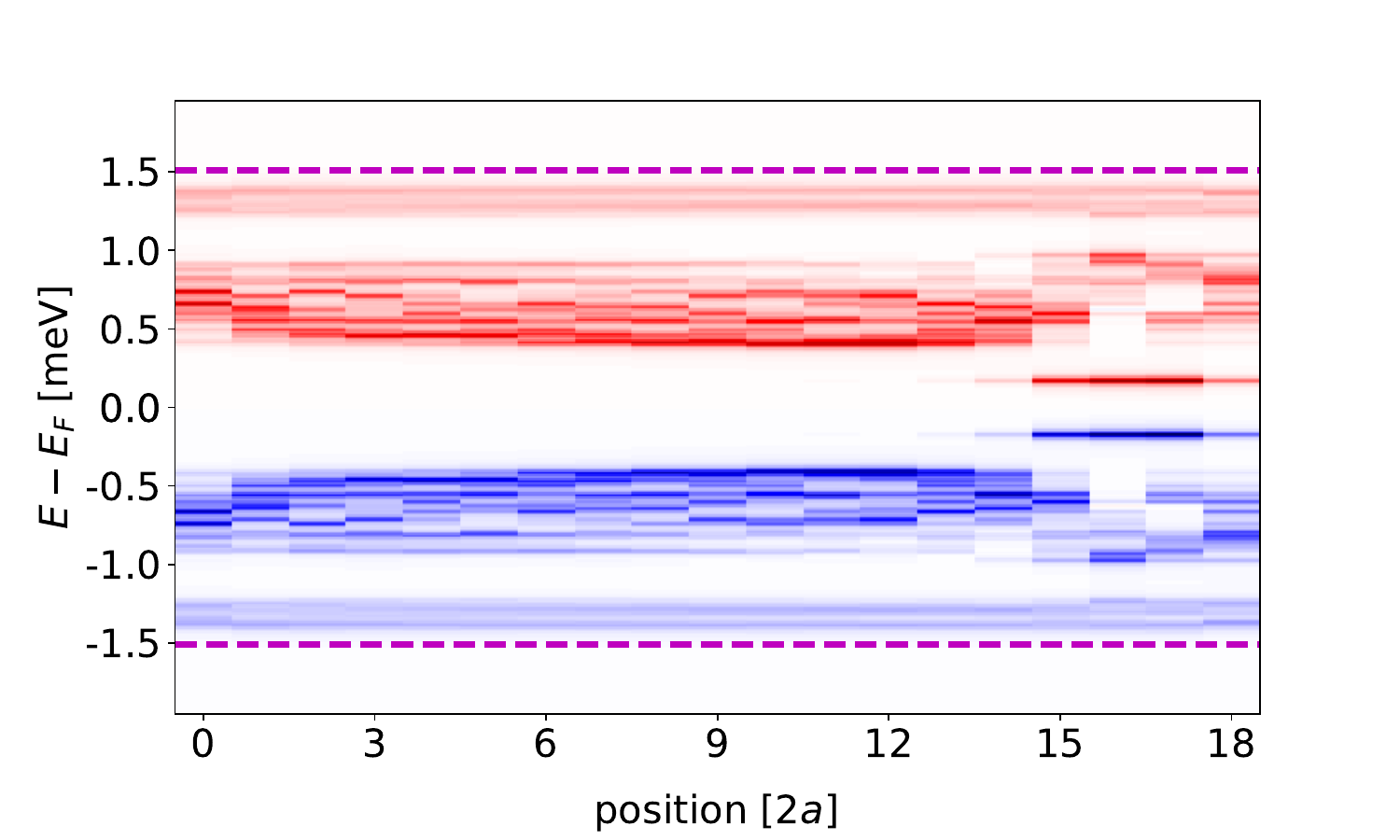}
    \put (0,50) {\normalsize (d)}
    \end{overpic}
\caption{\label{fig_comp1} The combination of spin spirals with different spiraling angles.
    (a) Local DOS for a spin spiral starting with a 175$^\circ$ spiraling angle from the left
     and with a $100^\circ$ spiraling angle from the right encountering at the sixth site (site number 5) where the spin points to the $z$ direction.
    (b) the Local CDOS for the same spin spiral.
    (c) and (d) represents the case when the spirals meet at the sixteenth site. }
\end{figure*}

In order to explore the basic elements of a simple braiding protocol,
we performed a series of computer experiments where artificial chains were created with two different spiraling angles on either side of the Fe$_{19}$ chain on Au/Nb(110).
From one side a $175^\circ$ spiral was launched which represents a topologically trivial  state (see paper I)
and hence no MZMs are expected to appear at the ends of these spirals.
From the opposite side a $100^\circ$ spiral (representing a topological state with MZMs)
was initiated towards the middle.
At the connection site the spin points along the $z$ axis, which is an element of both spirals. 
The results are presented in Fig.~\ref{fig_comp1} for the combination of these spin spirals.
It can be seen that in the left section of the chains, where no MZMs are expected,
we do not find any, and in the right part where MZMs are supposed to appear, we do find them, as it is presented in the first row of Fig.~\ref{fig_comp1}.
This calculation proves that MZMs are indeed localized at the boundaries of the topological region, however, it also has a practical consequence since it enables to shift MZMs along a chain by rotating the magnetic moments in various sections of the chain in unison.
Looking at the figures, one can see two hybridized states inside the minigap around zero energy, when the $175^\circ$ spiral part is left long, and the $100^\circ$ spiral part left short (Fig.~\ref{fig_comp1}c and d). These states  originate from the two extremely overlapping MZMs of the short part of the spiral.
It is rather \red{interesting} to observe, as the size of the minigap is changing from atom to atom and the gap from being trivial to topological as one crosses over from one section of the spiral to the other (see Fig.~\ref{fig_comp1} first row). \red{It also shows that the CDOS can be used quite reliably to indicate band inversion and the topological nature of the minigap.}

\red{Even though a single nanowire hosting a pair of MZMs
is insufficient to implement a topological qubit}, the observed shift of MZMs represents a significant first step
in the direction of the experimental realization of braiding and topological quantum gates.
If MZMs are constrained to move along a one-dimensional chain,
they must pass through one another in order to be swapped. 
However, one may imagine a network of such nanowires\cite{Beenakker2020} (a tri-junction at the very least), where the topological nature can be changed locally via a site-dependent rotation of magnetic moments.
As we demonstrated here, such local rotations allow to shift MZMs through the network without passing each other, as they are localized at the ends of the topological part of the spin spirals. On the experimental side, similar local rotations of individual magnetic moments were achieved recently by electron spin resonance techniques paving the way for the implementation of topological quantum gates.
\red{A similar effect was also studied in Ref.~\onlinecite{Marra2022b}
where MZMs can slide along the wire by applying a rotating magnetic field.
While a more detailed theoretical description about the construction of Majorana quantum gates based on ESR-STM techniques can be found in Ref.~\onlinecite{Bedow2023}.
The results presented in our work gives further 
information for the construction of Majorana gates described in Ref.~\onlinecite{Bedow2023}
by providing quantitative predictions for spiraling angles of different topological domains that can support 
the experimental realization.}

\section{Local manipulation of individual spins and the formation of Quasi-Majorana Bound States\label{sec:QMBS}}
\begin{figure}
    \begin{overpic}[width=0.45\textwidth]{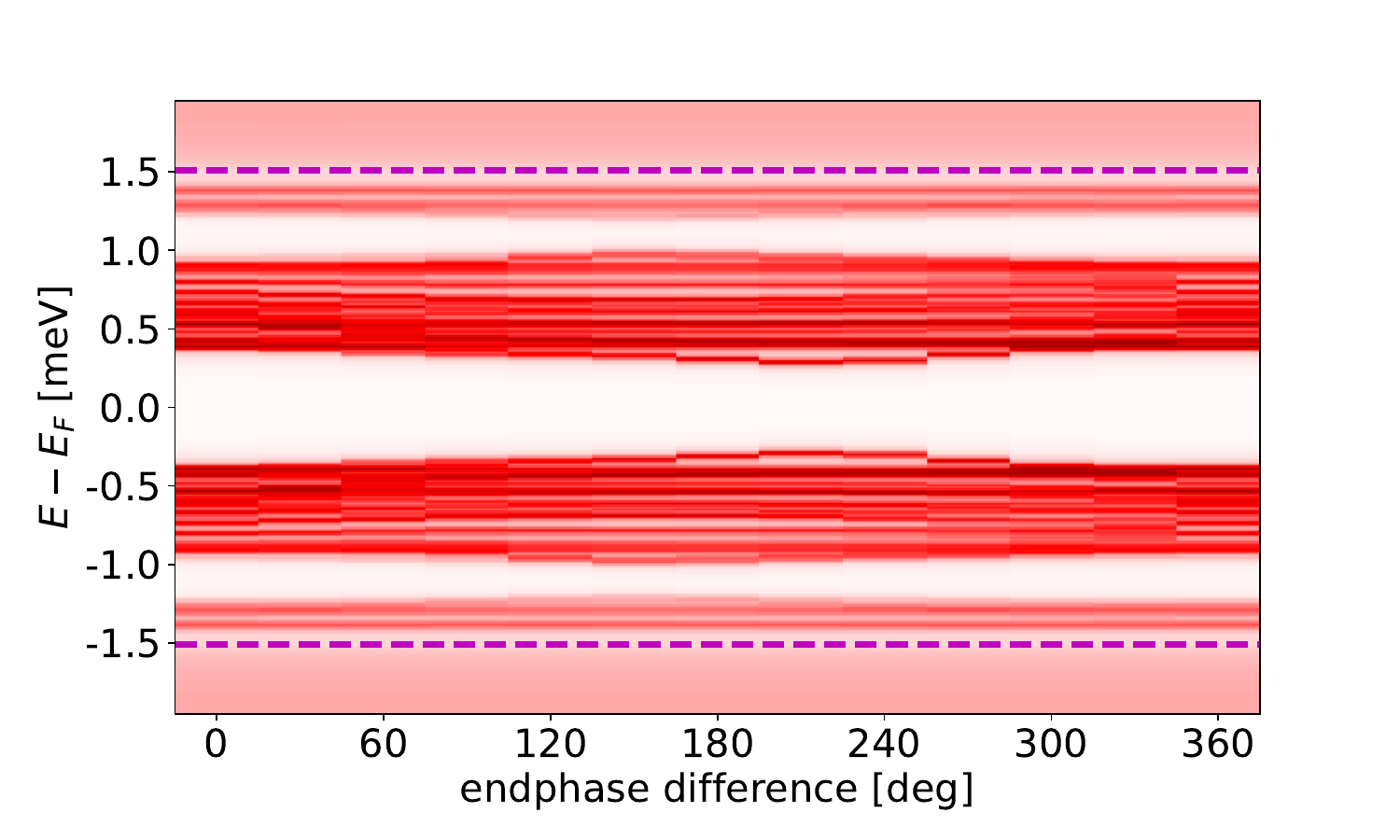}
    \put (0,50) {\normalsize (a)}
    \end{overpic}
    \begin{overpic}[width=0.45\textwidth]{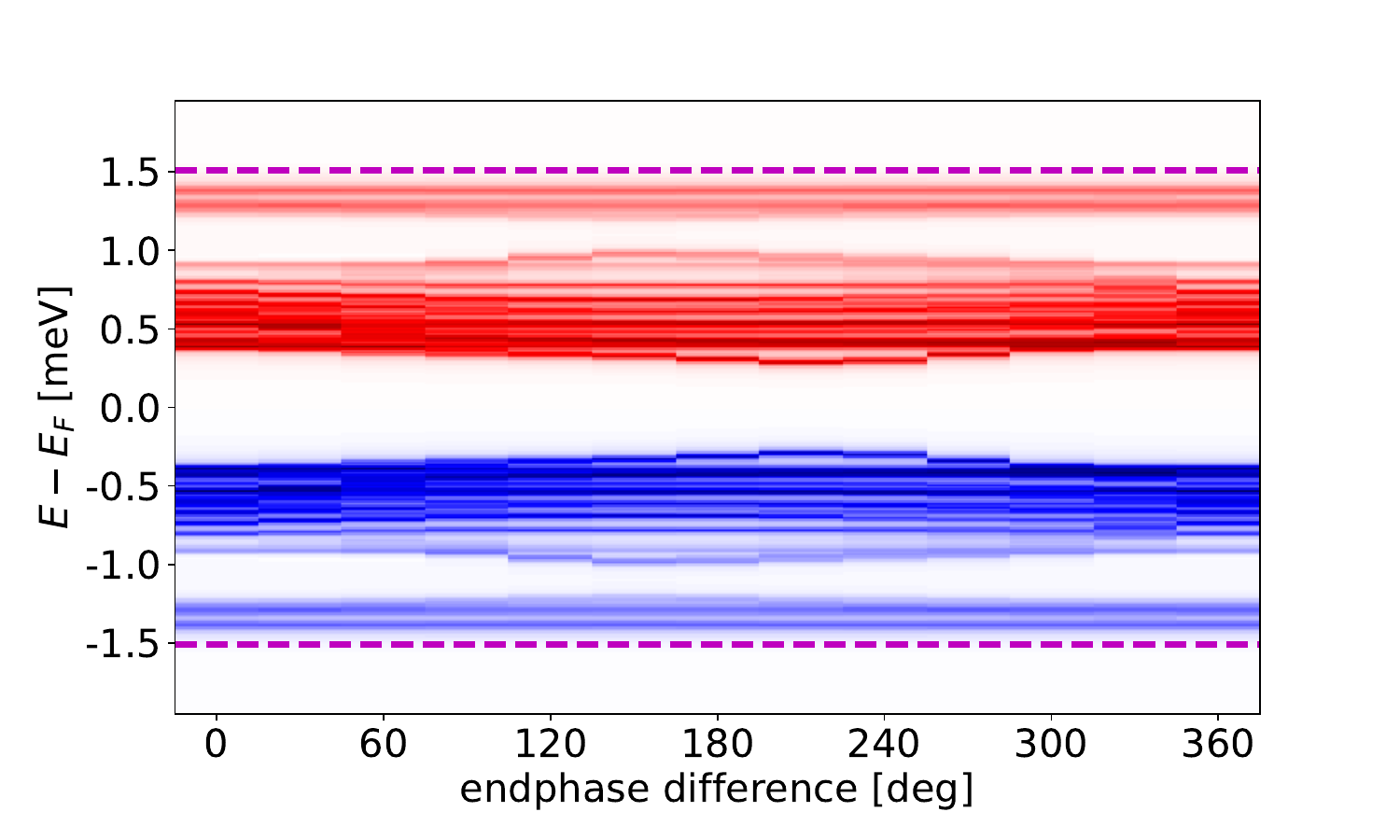}
    \put (0,50) {\normalsize (b)}
    \end{overpic}
\caption{\label{fig:phase_diff2}Topologically trivial spin spiral state with different end-phases. The total DOS~(a) and CDOS~(b) as a function of end-phase shift for a nanowire in the topologically trivial spin spiral state with spiraling angle 175$^\circ$. At the ends of the nanowire, the directions of the spins were rotated by the same amount.}
\end{figure}

\begin{figure}
	\centering
    \begin{overpic}[width=0.45\textwidth]{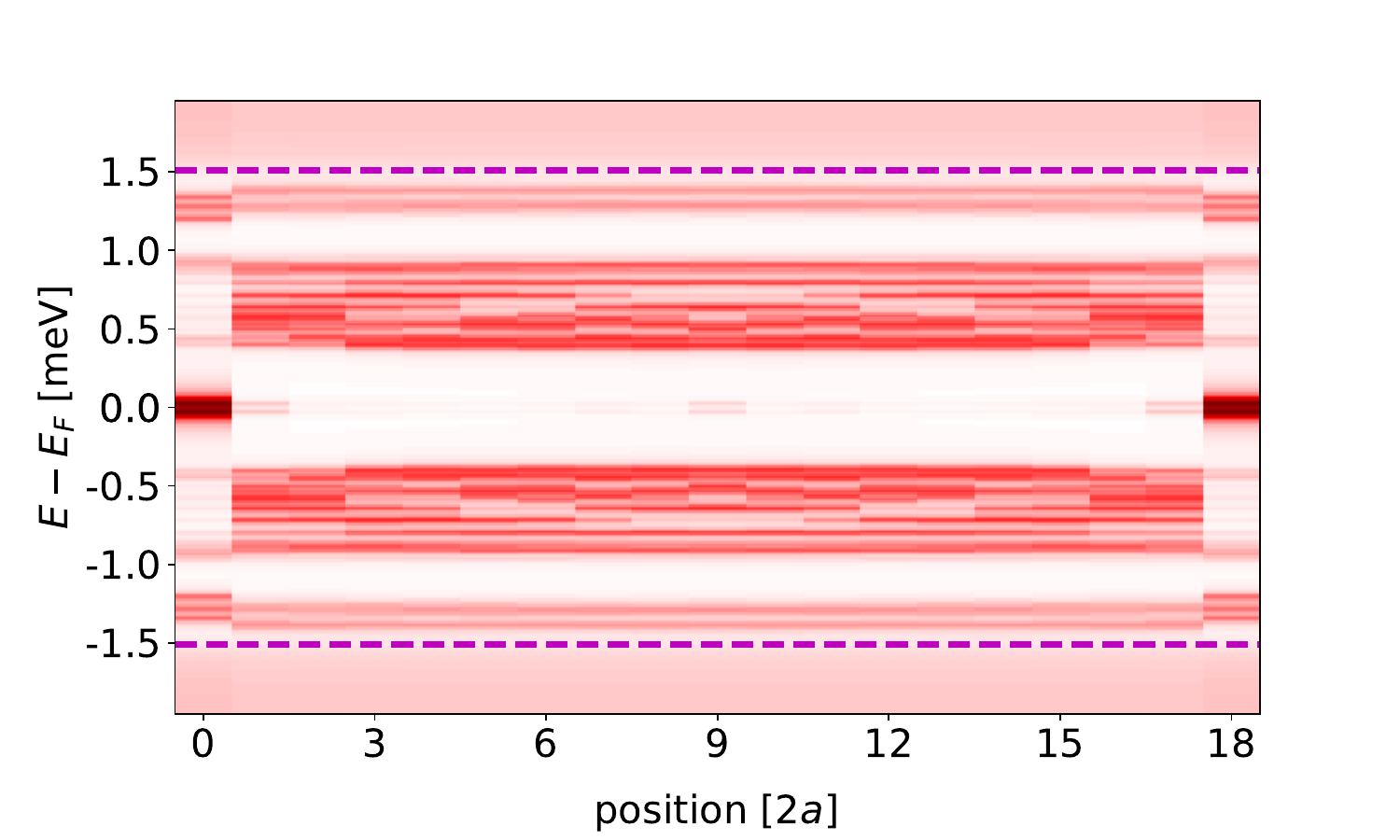}
    \put (0,50) {\normalsize (a)}
    \end{overpic}
    \begin{overpic}[width=0.45\textwidth]{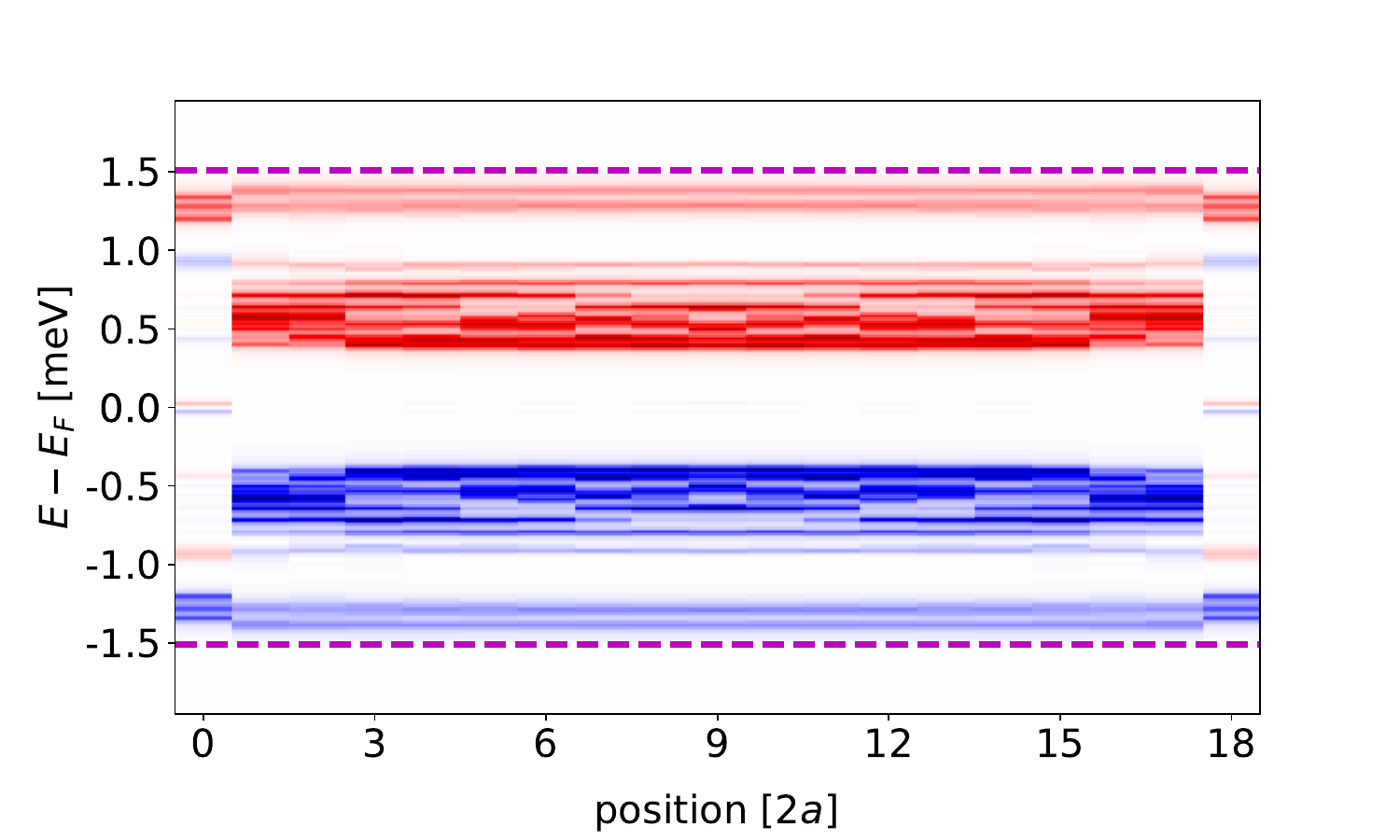}
    \put (0,50) {\normalsize (b)}
    \end{overpic}
\caption{\label{fig:quasimajorana}Topologically trivial Quasi-Majorana Bound States. The site-resolved Local DOS~(a) and Local CDOS~(b) for a chain where the magnetic moment of edge atoms has been artificially reduced by 2.5\% compared to the SCF result in a topologically trivial spin spiral state with $175^\circ$ spiraling angle.}
\end{figure}

We now turn our attention to the formation of 
topologically trivial zero energy states at the edges of the nanowire
that we shall call as Quasi-Majorana Bound States (QMBS).
We start our investigations with selecting  a spin spiral with 175$^\circ$ spiraling angle in the non-topological domain. 
In Ref.~\onlinecite{Kyungwha2023} it was shown that a single magnetic impurity can host a zero energy state by imposing
certain canting angles. Here one should remember how Shiba bands are formed.
If the chain contains magnetic adatoms sufficiently far away from each other, the system behaves
as a set of non-interacting single impurities with the same YSR states. When the magnetic adatoms are brought closer to each other,
the individual YSR states start to hybridize creating the so-called Shiba bands. 
Therefore, it is quite clear that in a dilute chain, topologically trivial zero energy states might appear as demonstrated in Ref.~\onlinecite{Kster2022}.
In fact, one may speculate that phase shifts occur at the ends of most 
chains due to locally different potentials, which may create QMBS.
The appearance of such QMBS can be tested within our method, by imposing a series of simultaneous phase shifts at both
ends of the nanowire in topologically trivial spin spirals. This is presented in Fig.~\ref{fig:phase_diff2}, where 
the phase shifts between 0$^\circ$ and 360$^\circ$ were applied simultaneously at the edge atoms.
Both the LDOS and LCDOS plots indicate that a local perturbation of the spin directions
at the ends of the 2a-[100] Fe$_{19}$ nanowire can not cause topologically trivial zero energy states at the edges.
Rotating around other spins within the iron nanowire leads to the same conclusion.
Therefore, we can conclude that the appearance of QMBS due to the misalignment of spins is highly unlikely in this particular nanowire.

However, local perturbations at the edges of the wire can affect not only the canting angles of the magnetic moment but also its magnitude.
To account for this possible mechanism, we artificially scaled the local exchange field simultaneously at both ends of the nanowire.
Quite generally, the scaling of the local exchange field does shift the energy of the YSR states
as it is also illustrated by the tight-binding approach in Appendix~A/2.
However, these calculations revealed a strong sensitivity of the YSR states to changes in the local exchange field.
In fact, a reduction of only 2.5\% in the local magnetic moment leads to rather surprising changes at the edges.
As Fig.~\ref{fig:quasimajorana} illustrates, even this small change was adequate to induce zero energy edge states in spirals where normally no such states were found.
At the same time, one can clearly observe the lack of band inversion on the LCDOS plot of Fig.~\ref{fig:quasimajorana},
which clearly indicates that the observed zero energy edge states are indeed QMBS not showing the expected topological properties.
Although there were a number of studies about potential mechanisms causing QMBS\cite{Dibyendu2013,Liu2017, Penaranda2018, Moore2018, Pan2020, DasSarma2021} and their experimental features\cite{Vuik2019,Liu2021},
they focused on other proposals of MZM platforms (such as semiconductor wires) without presenting quantitative predictions.
However, we can now rule out the appearance of QMBS due to local phase shifts and more importantly, predict the accurate value of the necessary  change in the local exchange field that leads to the appearance of zero energy edge states which are useless for topological quantum computing.

\section{Summary \label{sec:summary}}

In this paper, with the aim of taking into account the microscopic complexity,
we have applied the first-principles-based KKR Green's function method\cite{Csire2015, csire2018relativistic, Saunderson2020b, Nyari2021, saunderson2022full} to describe the superconducting ground state of nanowires with artificial spin spirals.
Beyond the actual calculations, our aim was to show, that the entire approach is an effective tool for inferring the existence of Majorana Zero Modes (MZMs) and investigating their characteristics in real materials.  
We were able to show the surprising stability of Majorana Zero Modes against fluctuations in the direction of magnetization of the atoms in the chain. 
Our calculations also revealed the strong sensitivity of the energy of the Yu-Shiba-Rusinov states to the changes in the local exchange field. An important consequence of this behavior is the appearance of non-topological Quasi-Majorana Bound States at the edges of the nanowire in the non-topological domain of spiraling angles. Their appearance is so defying, that to identify these states as QMBS, it was necessary to possess a computational tool, which could reliably differentiate between spirals with topological and non-topological energy spectra. 
It is quite fascinating that even a slight change in the local spin momenta at the edges of the nanowire can be sufficient to induce such zero energy edge states. Such imitative states have the potential to mimic many features of Majorana Zero Modes which could lead to inaccurate experimental conclusions. 
In addition, we identified a mechanism that poses a potential threat to Majorana Zero Modes localized at the edges of the nanowire. A jump in the local magnetization direction at some site, described by an additional phase shift on the spin spiral states can cause topological fragmentation and, therefore, Majorana Zero Modes may appear in the internal region of the nanowire and hybridize with the topological edge states.
Moreover, we have shown that the combination of spirals with different spiraling angles allows
the shift of Majorana Zero Modes. Since the recent developments of electron-spin-resonance
techniques allow such manipulations, these results open new avenues to engineer and implement braiding protocols of quantum gates in magnetic nanowires.

\section*{Acknowledgements}

B. Ny., A.L. L.Sz. and B.U. acknowledge financial support by the National Research, Development, and Innovation Office (NRDI Office) of Hungary under Project Nos. FK124100, K131938 and K142652. B. Ny. and L.Sz. acknowledge support by the Ministry of Culture and Innovation  and the NRDI Office within the Quantum Information National Laboratory of Hungary (Grant No. 2022-2.1.1-NL-2022-00004). B. Ny. acknowledges the support by the ÚNKP-21-3 New National Excellence Program of the Ministry of Culture and Innovation from the source of the NRDI Fund. The authors acknowledge KIFÜ for awarding us access to resources based in Hungary.

\section*{Appendix}

\subsection{Toy model of spin spirals in a nanowire}

The most simple microscopic model, that accounts for
the superconducting ground state of a spin spiral nanowire 
\red{with the same magnetic configuration as assumed in the realistic iron chain}),
is a tight-binding model with only nearest-neighbor interaction
and local on-site pairing interaction.
This can be described by the following mean-field Hamiltonian:
\begin{equation}
\begin{split}
    H &= \sum_{i,\alpha} \left( t c^\dagger_{i,\alpha} c_{i+1,\alpha}
    + \textrm{H.c.} \right)
    - \sum_{i,\alpha} \mu c^\dagger_{i,\alpha} c_{i,\alpha} \\
    & + \sum_{i,\alpha, \beta} \left( \Vec{B}_i \Vec{\sigma} \right)_{\alpha \beta}
       c^\dagger_{i,\alpha} c_{i,\beta}
      + \sum_i  \left( \Delta c^\dagger_{i,\uparrow} c^\dagger_{i,\downarrow}
       + \textrm{H.c.} \right)
\end{split}
\end{equation}
where $c^\dagger_{i,\alpha}$ and $c_{i,\alpha}$ denote the electron creation and annihilation operators
on site $i$ with spin $\alpha$, respectively. $\mu$ is the
chemical potential, and $\Delta$ is the $s$-wave type superconducting pairing potential accounting
for the interaction with the underlying conventional superconducting substrate.
There are several problems with such oversimplified models:
the proximity-induced superconducting pairing is introduced with a parameter and not via an interaction with the superconducting substrate
(causing inaccurate localization length of MZMs); both the multi-orbital nature and spin-orbit coupling are neglected, etc.
Despite its shortcomings, it can nevertheless grasp many important physical phenomena.
\red{The topological phases stemming from the interplay between spin-singlet superconductivity and a number of magnetic textures 
have been studied in Ref.~\onlinecite{steffensen2022topological}.}

\begin{figure*}
 \centering
 \includegraphics[width=0.3\textwidth]{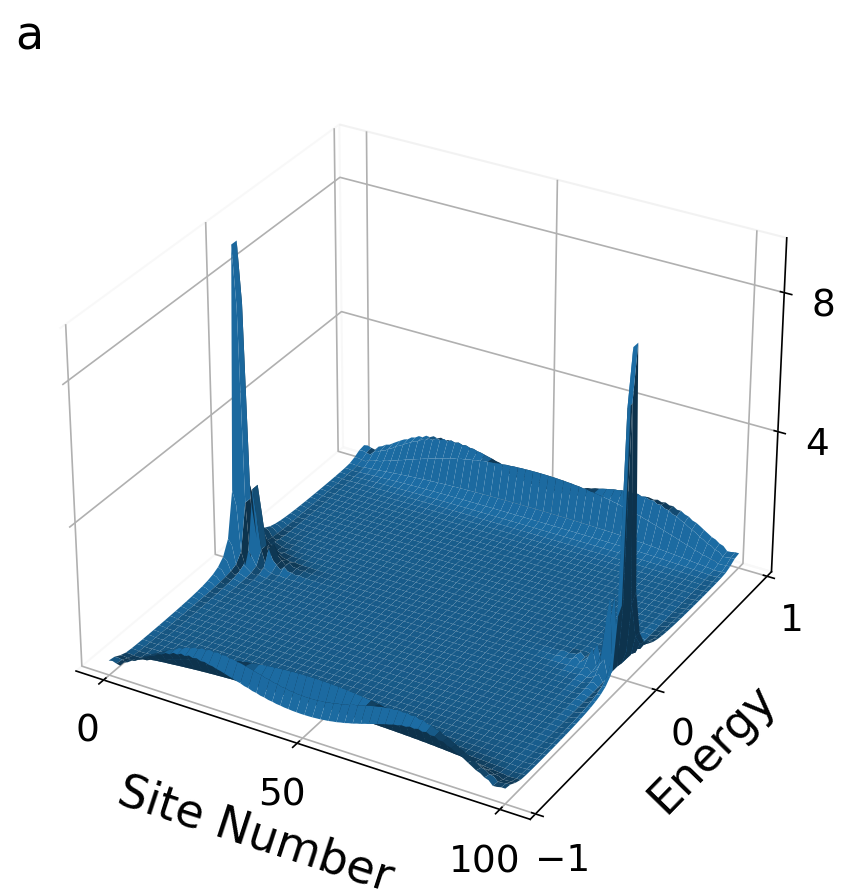}
 \includegraphics[width=0.3\textwidth]{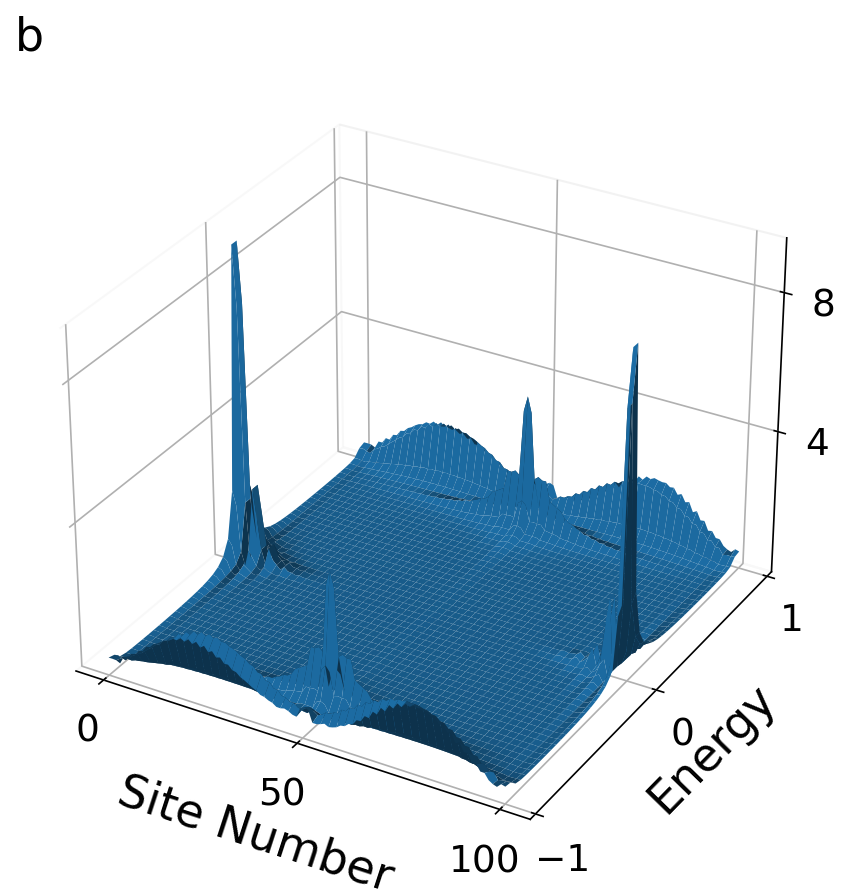}
 \includegraphics[width=0.3\textwidth]{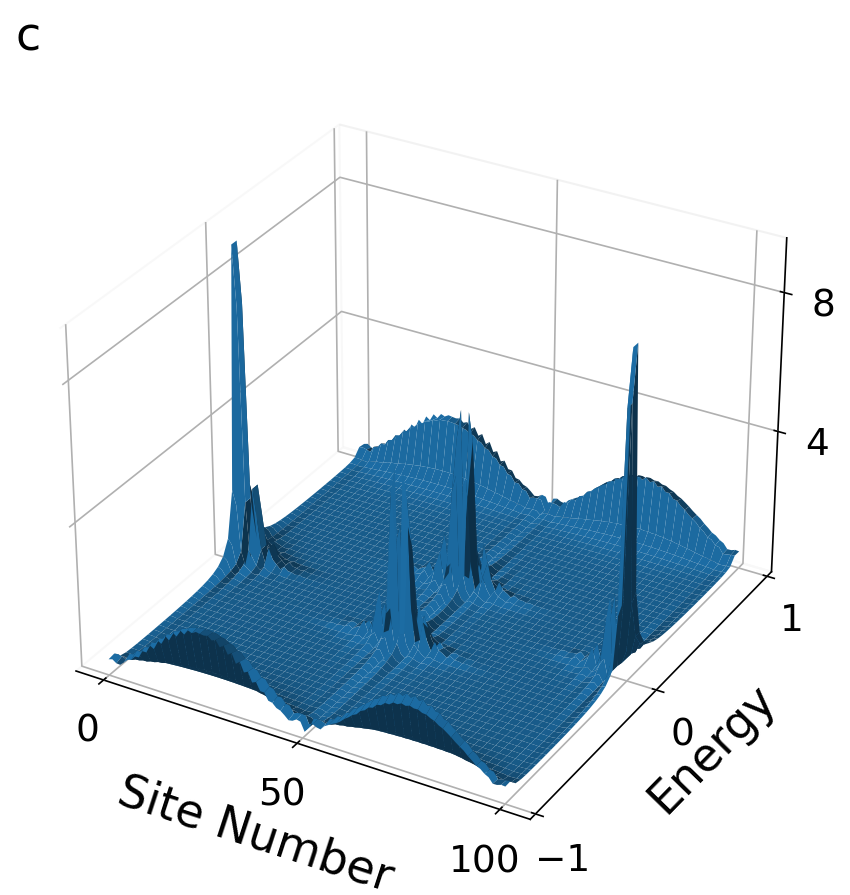}\\
 \includegraphics[width=0.3\textwidth]{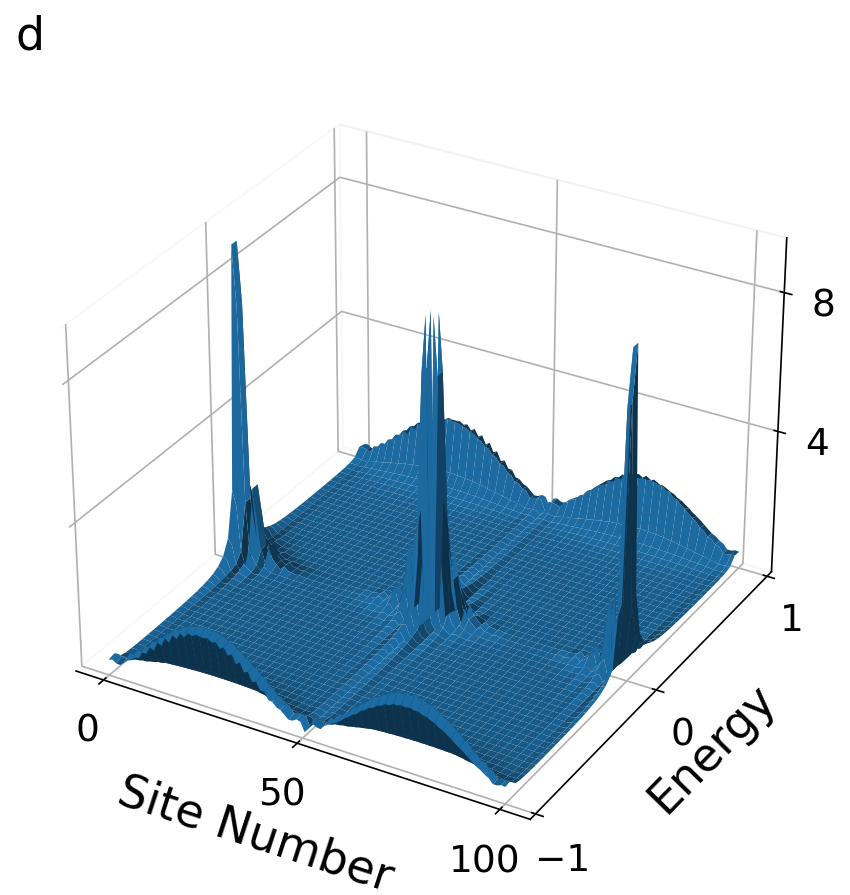}
 \includegraphics[width=0.3\textwidth]{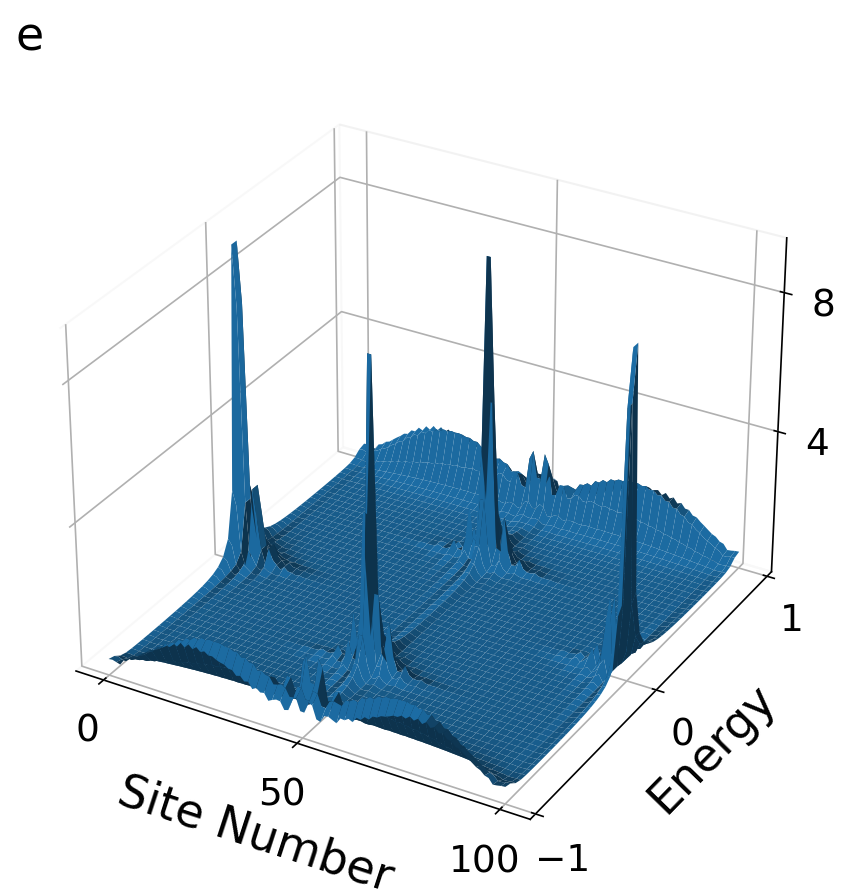}
 \includegraphics[width=0.3\textwidth]{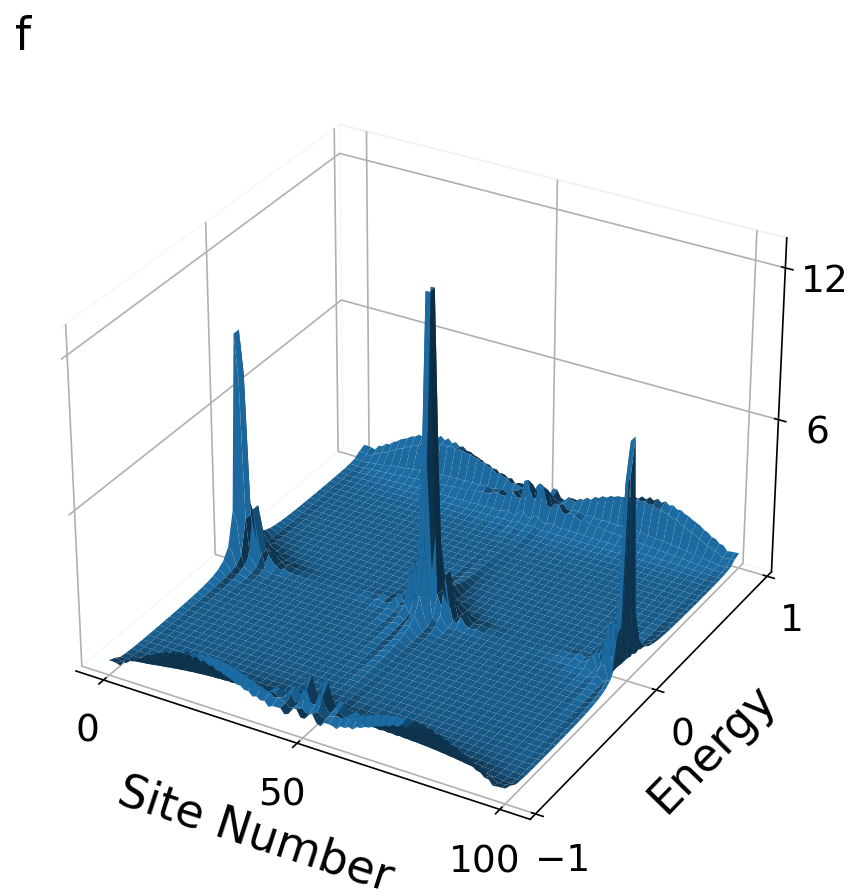}
		\caption{\label{fig:tb_model} The formation of topological fragmentation.
         Local DOS from the tight-binding model for the topological chain with 100 sites.
        (a) MZMs at the ends of the chain ("homogeneous" spin spiral);
        (b-d) manipulating the local exchange field ($\vec B_{n=50}$) in the middle of the chain by the scaling factor of
        (b) $0$,        
        (c) $10$,
        (d) $40$;     
        (e-f) introducing phase difference on the spin spiral in the middle of the chain as
        (e) $\Delta \phi=120^\circ$ and     
        (f) $\Delta \phi=155^\circ$.}
\end{figure*}

\begin{figure*}
 \centering
 \includegraphics[width=0.3\textwidth]{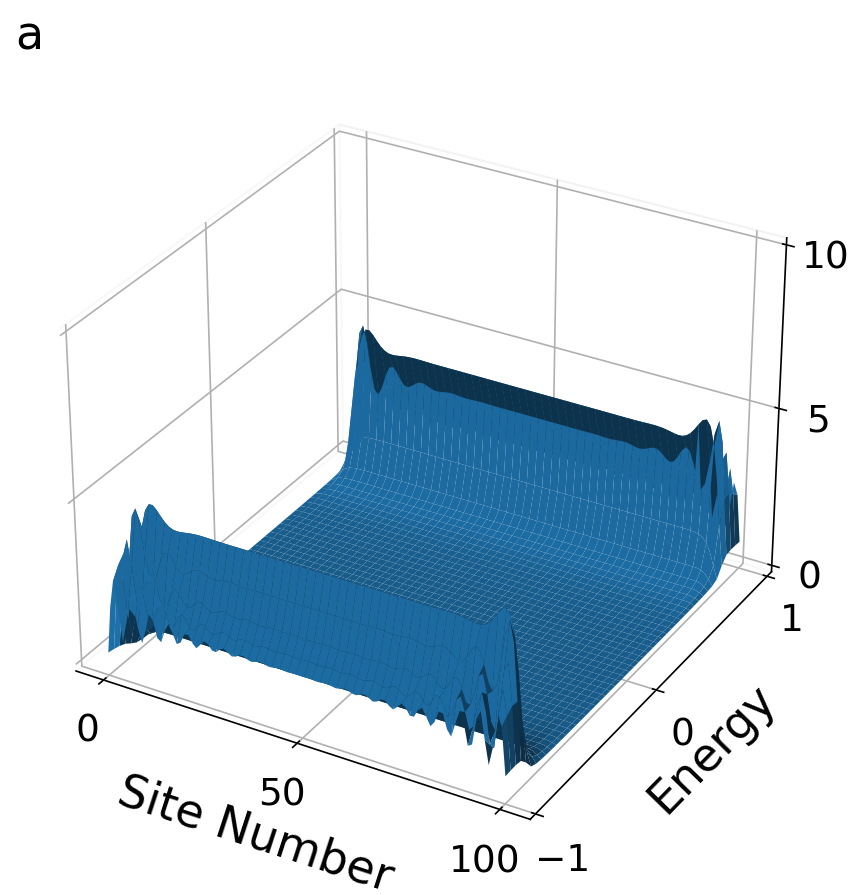}
 \includegraphics[width=0.3\textwidth]{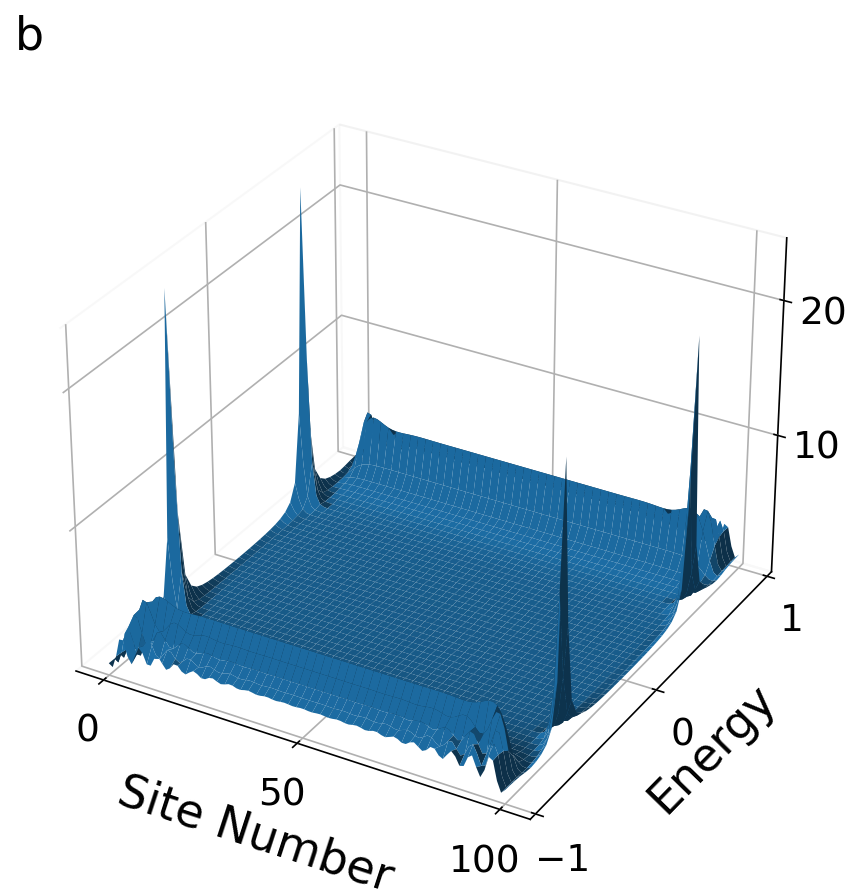}
 \includegraphics[width=0.3\textwidth]{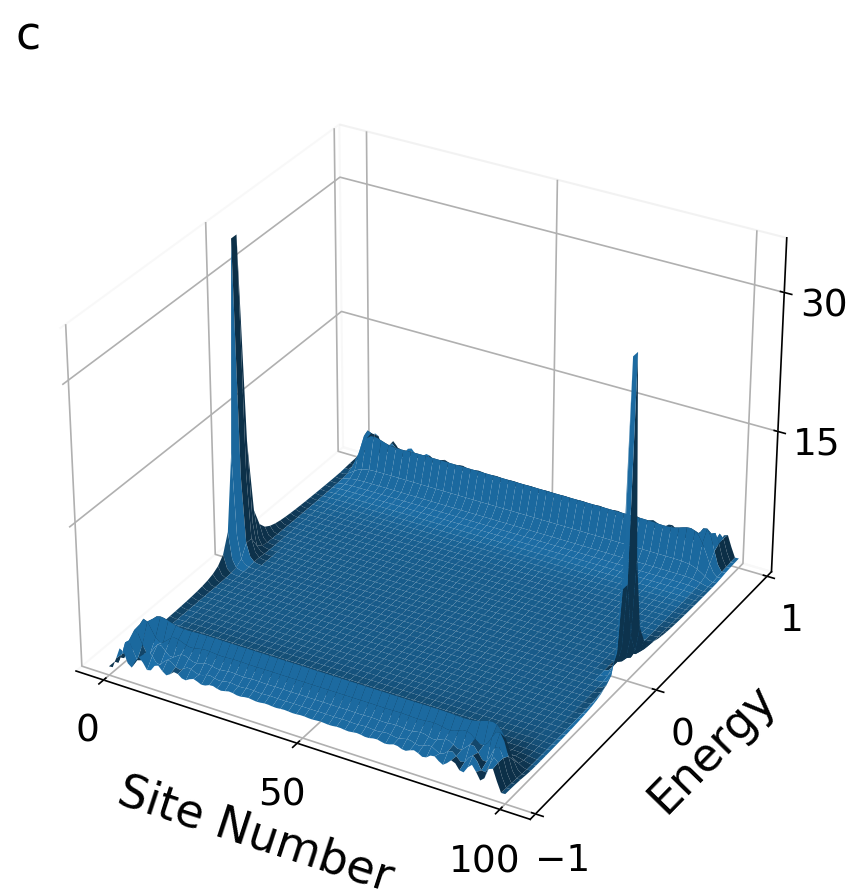}
		\caption{\label{fig:tb_model2} The formation of Quasi-Majorana-Bound States.
        Local DOS from the tight-binding model for the non-topological chain in
        (a) the topologically trivial phase;
        (b-c) with manipulating the local exchange field at the ends of the chain by the scaling factor of
        (b) 1.8, and        
        (c) 2.12, respectively.}
\end{figure*}

\subsubsection{The formation of topological fragmentation}

First, we tune the tight-binding model into a topological state.
This is achieved by taking
$t=4$, $\Delta=1$, $\mu=3$, spiraling angle $\pi/2$, $|\Vec{B}|=2.5$ on a chain containing 100 sites.
We shall draw consequences based on the local density of states (LDOS) of the quasi-particles.
In Fig.~\ref{fig:tb_model}(a) one can observe the localized Majorana Zero Modes at the ends of the chain.
One should keep in mind that in the topological phase, Majorana operators from different sites are paired together\cite{kitaevchain}.
As a first attempt, we try to create a fragmented chain from the perspective of Majorana Zero Modes
by changing the amplitude of the local exchange field in the middle of the chain.
In Fig.~\ref{fig:tb_model}(b) we show the result with zero local exchange field
suggesting a fragmented chain.
The most striking feature is the appearance of a finite energy state which stems from the interaction of two MZMs in the middle of the chain.
In Fig.~\ref{fig:tb_model}(c-d) we scaled up the local exchange field leading to an even more interesting effect:
we have achieved a perfect fragmentation of the chain inducing two MZMs in the middle of the chain without interacting with each other.
The reasoning behind the interpretation of topological fragmentation is justified by the fact that further increasing the local exchange field preserves the observed zero energy state
which is not possible to achieve with spin spiral states belonging to the non-topological domain.

We consider the most interesting scenario in Fig.~\ref{fig:tb_model}(e-f): an additional phase is introduced in the middle of the chain. 
Surprisingly, the effect of this additional phase resembles the manipulation of the local exchange field.
We again observe a finite energy state appearing in the gap region, which
can be tuned into zero energy by changing the value of the additional phase.
Most importantly, in this model, a similar effect can not be observed in a non-topological chain. By other words, one can not push any state into the gap region by imposing phase differences in spin spiral states belonging to the non-topological domain of the TB parameters.
Hence, at around $155^\circ$ one may observe again a perfectly fragmented chain from the point of view of Majorana operators.
However, as we have seen with the \textit{ab initio} technique, such fragmentation may cause larger problems for MZMs in the case of shorter chains: 
the MZMs at the edges may hybridize with the (new) state in the middle and evolve into a pair of finite energy YSR states.
Additionally, we mention that since the host wasn't involved in this approach,
the Néel and Bloch-type spin spirals have exactly the same effects
suggesting that the concept of topological fragmentation is a rather general effect.

\subsubsection{The formation of Quasi-Majorana Bound states}

Here we address the possibility of the formation of non-topological zero energy edge states, based on the simple tight-binding model. First, we tune the model into the non-topological domain,
where MZMs are not expected, by taking the following parametrization:
$t=4$, $\Delta=1$, $\mu=3$, spiraling angle $\pi/2$, $|\Vec{B}|=2.5$.
As Fig.~\ref{fig:tb_model2}a displays the MZMs are indeed absent
at the edges of the chain.
Then we assume a local field effect at the edges of the chain which
scales the local exchange field.
By increasing this local exchange field at the edges, one can observe
(see Fig.~\ref{fig:tb_model2}b)
the appearance of in-gap YSR states.
In fact, one can find a scaling value, where these in-gap states are exactly at zero energy and, therefore,
mimic the behavior of MZMs. These states can be called Quasi-Majorana-Bound States (QMBS, see also Sec.~\ref{sec:QMBS}).
An interesting aspect of such a model calculation is that the model reacts differently to the scaling of the local exchange field in the topological and non-topological domain.
In topological spin spirals
the increase in the local exchange field is pushing the states toward zero energy, but above a critical value, these states remain at zero. Also, the scaling factor that is needed to reach zero energy is quite high, (around a factor of 40).
In contrast, in the non-topological spirals, one can continuously shift the in-gap states by the scaling of the local exchange field.
Moreover, the exact energy position of the in-gap states changes quite sensitively in response to the local scaling, which 
is again in contrast to the topological case as one can observe by looking  at the scaling factors in the captions of Figs.~\ref{fig:tb_model}-\ref{fig:tb_model2}.


\subsection{Computational details}
The details of the self-consistent calculations were summarized in Paper I, here we focus on some minor differences.
As we mentioned in the introduction, we used a frozen potential approach, and we performed self-consistent calculations only in the ferromagnetic state. This way, we also neglected the somewhat arbitrary definition of the induced moments during the rotation of the strong magnetic moments. Compared to control calculations on select systems, this caused very small effect, and we pointed out the differences  in Sec.~\ref{sec:phase}. 
To have an accurate description of our system within these approximations, we used  the largest number of $\mathbf{k}$ points used in Paper I (1891) in the irreducible wedge of the Brillouin zone (BZ), for all types of calculations (self-consistent bulk, surface, impurity cluster calculations, and also for the non-self-consistent calculations in the superconducting state).

The superconducting DOS and CDOS were calculated by a single-shot calculation by solving the KSDBdG equation with the effective pair interaction $\Delta=1.51$~meV\cite{Beck2021} in the Nb layers\cite{Nyari2021},  obtained from the experimental band gap  A sufficient energy resolution of the LDOS in the superconducting gap is acquired by considering 301 energy points between $\pm 1.95$~meV with an imaginary part of 13.6~$\mu$eV related to the smearing of the resulting LDOS.

\bibliography{Nyari_main}

\end{document}